\documentclass[a4paper,12pt]{article}
\usepackage[utf8]{inputenc}
\usepackage{amsmath}
\usepackage{amssymb}
\usepackage{bbold}
\usepackage{amsfonts,bm}
\usepackage{geometry}
\usepackage{fullpage}
\usepackage{graphicx}
\usepackage{caption}
\usepackage{todonotes}
\usepackage{multicol}
\usepackage{subcaption}
\usepackage{slashed}
\usepackage{float}
\usepackage{color,xcolor}
\usepackage{epstopdf}
\usepackage{units}
\usepackage[normalem]{ulem} 
\usepackage{jheppub} 
\usepackage{bm,paralist,xspace,url,relsize}
\usepackage[]{hyperref}
\newcommand{\ldec}{\ensuremath{l_{\text{decay}}}\xspace}
\newcommand{\lmin}{\ensuremath{l_{\text{min}}}\xspace}
\newcommand{\lmax}{\ensuremath{l_{\text{max}}}\xspace}
\newcommand{\ldet}{\ensuremath{l_{\text{det}}}\xspace}
\newcommand{\nev}{\ensuremath{N_{\text{events}}}\xspace}

\newcommand{\pdec}{\ensuremath{P_{\text{decay}}}\xspace}
\newcommand{\egeom}{\ensuremath{\epsilon_{\text{geom}}}\xspace}
\newcommand{\brvis}{\ensuremath{\text{Br}_{\text{vis}}}\xspace}

\usepackage{multirow}
\graphicspath{{./plots/}}

\title{Searches for new physics at SND@LHC}
 \author[a]{Alexey~Boyarsky,}
 \author[a,b]{Oleksii~Mikulenko,}
 \author[a]{Maksym~Ovchynnikov,}
 \author[c]{and Lesya~Shchutska}
  \affiliation[a]{Instituut-Lorentz, Leiden University, Niels Bohrweg 2, 2333 CA Leiden, The Netherlands}
  \affiliation[b]{Department of Physics, Taras Shevchenko National University of Kyiv, 64 Volodymyrs’ka str., Kyiv 01601, Ukraine}
 \affiliation[c]{Institute of Physics, École Polytechnique Fédérale de Lausanne (EPFL), BSP 616, Rte de la Sorge, CH-1015 Lausanne, Switzerland}

\emailAdd{boyarsky@lorentz.leidenuniv.nl}
\emailAdd{mikulenko@lorentz.leidenuniv.nl}
\emailAdd{ovchynnikov@lorentz.leidenuniv.nl}
\emailAdd{lesya.shchutska@epfl.ch}
 
\date{}

\begin{document}
\abstract{SND@LHC is an approved experiment equipped to detect scatterings of neutrinos produced in the far-forward direction at the LHC, and aimed to measure their properties. In addition, the detector has a potential to search for new feebly interacting particles (FIPs) that may be produced in proton-proton collisions. In this paper, we discuss signatures of new physics at SND@LHC for two classes of particles: stable FIPs that may be detected via their scattering, and unstable FIPs that decay inside the detector. We estimate the sensitivity of SND@LHC to probe scatterings of leptophobic dark matter and decays of neutrino, scalar, and vector portal particles. Finally, we also compare and qualitatively analyze the potential of SND@LHC and FASER/FASER$\nu$ experiments for these searches.}
\maketitle

\section{Introduction and summary}
\label{sec:introduction}
The Standard Model of particle physics (SM) successfully describes experimental data from accelerators. However, it fails to explain three observational phenomena: neutrino oscillations, dark matter, and the baryon asymmetry of the Universe. This indicates that some new particles responsible for these phenomena may exist in nature. 

New particles that have tiny couplings with the SM particles (feebly interacting particles, or FIPs) may be searched by the Intensity frontier experiments, which operate with high rate of particle collisions. During several last years, many such experiments have been proposed: SHiP~\cite{Alekhin:2015byh,Anelli:2015pba,Bondarenko:2019yob},
CODEX-b~\cite{Gligorov:2017nwh},
MATHUSLA~\cite{Chou:2016lxi,Evans:2017lvd,Curtin:2018mvb,Bondarenko:2019yob}, FASER~\cite{Feng:2017uoz,Feng:2017vli}, SeaQuest~\cite{Berlin:2018pwi},
NA62~\cite{Mermod:2017ceo,CortinaGil:2017mqf,Drewes:2018gkc} and a number of other experiments (see~\cite{Beacham:2019nyx} for an overview).

One of the recent proposals is the Scattering and Neutrino Detector (SND@LHC) facility~\cite{Ahdida:2020evc,Loi_SND,TP}, which aims to study scatterings of high-energy SM neutrinos produced in the far-forward direction at the LHC in the ATLAS interaction point. This detector can also be used to search for new particles that scatter similarly to neutrinos, such as light dark matter~(LDM) particles that interact with the SM particles via portal mediators, playing the role of FIPs. 

Depending on the model, there are various signatures of new physics in processes with LDM or mediators. Examples are: LDM scattering off electrons or protons (an incomplete list of studies is~\cite{Ahdida:2020evc,Batell:2009di,Batell:2014yra,Batell:2018fqo,Batell:2021blf,deNiverville:2016rqh,Coloma:2015pih,Naaz:2020tgx,Dobrescu:2014ita,Soper:2014ska,Frugiuele:2017zvx,SHiP:2020noy,Breitbach:2021gvv}); rare neutrino scattering events induced by new physics, as the trident reaction $\nu + n \to p + \chi + l$ with missing transverse momentum
~\cite{Kelly:2019wow,Kelly:2020pcy}, or the process $\nu+ Z \to \nu +Z+l+\bar l$~\cite{Altmannshofer:2014pba}; an excess of charged current (CC) $\nu_{\tau}$ events, such as in models with neutrinophilic mediators coupled exclusively to $\tau$ lepton flavor~\cite{Kling:2020iar}.

Some of these signatures have been studied for SND@LHC and a similar facility FASER$\nu$~\cite{Abreu:2019yak,Abreu:2020ddv}, see~\cite{Ahdida:2020evc,Kling:2020iar,Batell:2021blf,Bakhti:2020vfq,Bahraminasr:2020ssz,Bakhti:2020szu}. In this work, we consider the signature of LDM scattering off nucleons, which has not been considered previously.\footnote{The study of the scattering off protons in~\cite{TP} is based on this work.} In addition, we estimate the sensitivity of SND@LHC to decays of mediators. 

Whether SND@LHC may probe currently unexplored parameter space depends on the coupling of the mediator to electrons and photons. If it is present, the model may be already constrained by experiments that search for missing energy/momentum, such as NA64~\cite{NA64:2019imj}, BaBar, Belle~\cite{Lees:2017lec}. They require FIPs only to be produced, and therefore, given the coupling $g$ to the SM particles, the expected number of events is proportional to $g^2$. SND@LHC, in its turn, requires the produced particle to scatter/decay, and therefore, the number of events scales as $g^4$. Missing energy experiments are more sensitive to small couplings, and SND@LHC may not be able to probe unconstrained parameter space if they are relevant. This is the case, for instance, for the dark photon mediator~\cite{TP}.
\vspace{10pt}

Let us look closer at the LDM scattering off nucleons. This scattering may be mimicked by neutral current~(NC) neutrino scattering events, and therefore, such a search is not background-free. Typically, to observe a signal over background, many LDM scattering events are required~\cite{Batell:2014yra,Coloma:2015pih,Batell:2018fqo,Naaz:2020tgx,Dobrescu:2014ita}. Under this condition, one can look for an excess of a signal over the numerous neutrino background, and in particular to distinguish events with LDM and neutrinos kinematically by comparing their reconstructed energy spectra. It would be therefore attractive to consider signatures that require less amount of events.

Particle scattering may occur elastically or inelastically, producing either an isolated proton or hadronic jets correspondingly. Probabilities of these two processes have a different dependence on the mediator mass $m$: elastic scattering becomes subdominant for large masses $m\gtrsim 1\text{ GeV}$, see Sec.~\ref{sec:elastic-inelastic}. This is the case for neutrino scattering, with the mediators being the $W$ and $Z$ bosons. As a result, a very few of elastic neutrino scattering events are expected at SND@LHC. For LDM that interacts via light $\mathcal{O}(\unit[1]{GeV})$ mediators, the situation is different, see Fig.~\ref{fig:elastic-to-dis-ratio}. 
Therefore, elastic scattering signature could provide good sensitivity in this part of the parameter space.

In case of large mass of the mediator $m\gtrsim\unit[1]{GeV}$, inelastic scattering dominates, leading to an excess of neutral current events. The expected number of such events at SND@LHC is not known precisely because of theoretical uncertainty in the calculation of the neutrino flux in the far-forward direction, which reaches a factor of few~\cite{Kling:2021gos}. However, the ratio of the number of NC and CC events is known with a good precision in the SM, see Sec.~\ref{sec:charged-neutral}. In presence of LDM, this ratio may increase over the experimental uncertainty in measurements of the NC/CC ratio at SND@LHC, which is about 10\%~\cite{TP}.

In Sec.~\ref{sec:leptophobic} we consider a mediator that does not couple to leptons -- the leptophobic portal, and estimate the sensitivity of SND@LHC to LDM scatterings off nucleons using the elastic/inelastic and NC/CC signatures.

There is one more way to distinguish LDM and neutrinos if mass $m_\chi$ of an LDM particle is sufficiently large. Namely, massive particles reach the SND@LHC detector volume later than neutrinos. As a result, the neutrino background can be eliminated using the time-of-flight measurements. The time resolution of SND@LHC is $\unit[200]{ps}$~\cite{TP}, and after imposing the timing cut it would be possible to separate an event with LDM from neutrino scattering events for gamma-factors $\gamma_{\chi}<100$. However, particles flying in the far-forward direction at the LHC have large energies $E_{\chi}\simeq \unit[1]{TeV}$, and therefore, their mass should be of order of $m_{\chi}\simeq \unit[10]{GeV}$. Such massive FIPs are typically poorly produced and have suppressed scattering probability (at least, this is the case for the LDM interacting via dark photon and leptophobic portal, see Sec.~\ref{sec:scatterings}).

\vspace{10pt}

Let us now consider the signature of decays of mediators. Although SND@LHC is constructed to probe neutrino scatterings, it may also be capable of searching for decays of FIPs, for instance scattering mediators (as we argue in Sec.~\ref{sec:decays}). It is attractive to probe the parameter space simultaneously by scatterings of LDM and decays of mediators. However, for the given coupling $g$, the decay length is typically much shorter than the scattering length, see Appendix~\ref{app:scattering-vs-decays}. As a result, for large couplings, that are required to see scatterings, the decay length is microscopic, and mediators decay before reaching the detector. It may be still possible though to probe large couplings via scatterings of LDM and smaller couplings via decays of the mediator.

A clear background-free signature may be decays of a FIP into a di-lepton pair, $V \to l l'/l l' \nu$, as scatterings of neutrinos produce at most one lepton.\footnote{The di-muon events may be produced by the scattering of photons in the detector. However, the photons occur in scattering of neutrinos, and apart from the di-muon pair there would be a lot of other tracks.} For the decays of FIPs into a lepton and a meson, or into a pair of mesons there is a background that comes from the neutrino deep inelastic CC- and NC-scatterings correspondingly.
However, decay products typically carry large energies $E\gtrsim \unit[100]{GeV}$ and, therefore, can be distinguished from (inelastic) neutrino scatterings with such large energy transfers as the latter typically produce a lot of hadrons. Therefore, we believe that the mentioned background may be rejected. This question requires an additional study.

We estimate the sensitivity to the mentioned decays of scalar, neutrino and vector portals, see Sec.~\ref{sec:portals}.

In Sec.~\ref{sec:discussion}, we compare the potential of SND@LHC with FASER/FASER$\nu$ experiments to probe decays and scatterings, discussing both abilities of their detectors and the effect of their placements with respect to the beam axis (FASER is on-axis, while SND@LHC is slightly off-axis).
        
\section{SND@LHC experiment}
\label{sec:snd-experiment}
\begin{figure}[h!]
    \centering
    \includegraphics[width = 0.9\textwidth]{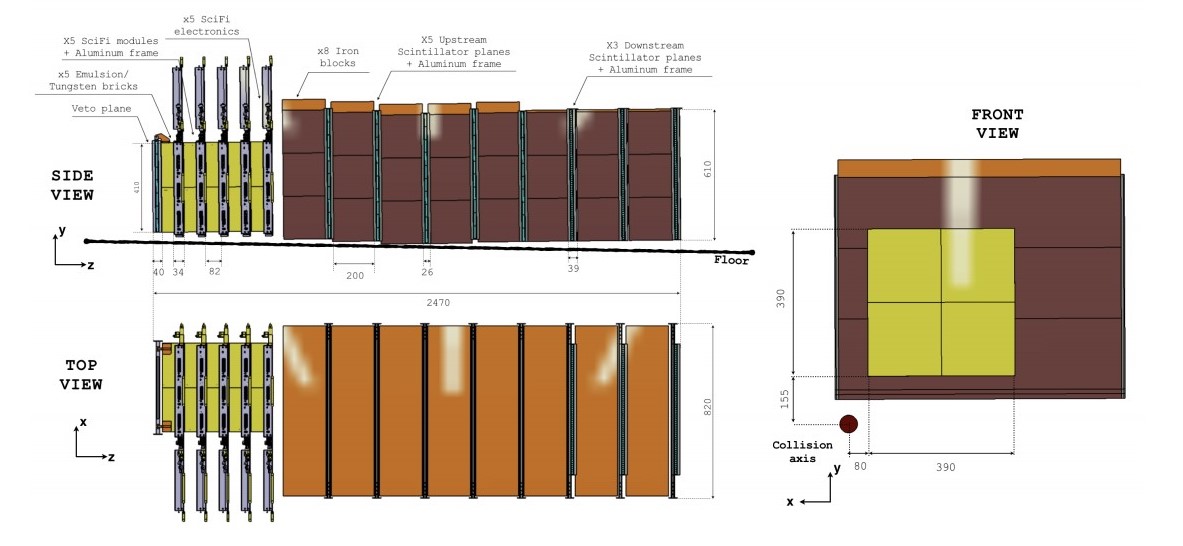}
    \caption{Overview of the SND@LHC detector facility: the side view (on the left) and the front view (on the right). The figure is taken from~\cite{TP}.}
    \label{fig:snd}
\end{figure}
SND@LHC facility is planned to be installed in the TI18 tunnel at the distance of 480~m from the ATLAS interaction point along the beam collision axis. The SND@LHC detector consists of the target region followed by the muon system, see Fig.~\ref{fig:snd}. The pseudorapidity range covered by the target will be $7.2<\eta<8.6$, in which $\nu_e$, $\nu_\tau$ are produced in decays of heavy mesons, with an additional component of muon neutrinos originated from decays of pions and kaons. The actual angular position of the target is $(\theta_x, \theta_y) \in [0.17, 0.98]\times [0.32, 1.14]\,\unit{mrad^2}$.

The target has brick structure: bricks of emulsion cloud chambers (ECC) followed by Scintillating Fibre (SciFi) plates. Each of five emulsion bricks consists of 60 emulsion films interleaved with 59 tungsten plates of 1 mm thickness, which serve as target. The total scattering length of the emulsion bricks is $\unit[29.5]{cm}$, which corresponds to 84 radiation lengths~($X_0$), and the total target length of about $\unit[40]{cm}$. The ECC provide micrometric accuracy that allows one to measure accurately tracks of charged particles, and reconstruct vertices of neutrino interactions and any other event that deposit their energy within one emulsion brick. The bricks will be replaced every six month after collecting $\unit[25]{fb^{-1}}$ of data.

SciFi measures the energy of electromagnetic and hadronic showers, and provides spacial timing information for showers and single charged particles (with resolution of the order $\unit[50]{\mu m}$ and $\unit[100]{ps}$ correspondingly). In this way, it serves for disentangling the piled up events occurring in the emulsion and connecting tracks between emulsion walls.

The target is followed by the muon system, which consists of eight iron blocks interleaved with scintillator planes. It identifies muons as the most penetrating charged particles and measures energy of hadronic cascades. The muon detector in combination with SciFi works as a hadronic calorimeter with 9-11 interaction lengths. The efficiency of muon identification in CC $\nu_\mu$ event is found to be $69\%$, while in NC the probability that hadrons are not misidentified as muons is above $99\%$.

An important feature of SND@LHC is high neutrino type identification efficiency. The target construction allows track detection of charged leptons produced in primary interactions and subsequent decays. Electrons are identified by electromagnetic showers, muons are detected in the muon system, while $\tau$ leptons -- via a displaced decay vertex from the primary interaction thanks to micrometric resolution of the ECC.

There are two phases of the event reconstruction at SND@LHC~\cite{TP}. The first phase uses electronic detectors: events are reconstructed based on veto, the target tracker and the muon system. The second phase adopts the emulsion target, and the event reconstruction will be available six months after the exposure. It is crucial for identification of events with $\tau$-leptons, with the efficiency $\approx 50\%$. Moreover, it distinguishes EM showers induced by electrons from those from photons/$\pi^0$ with efficiency above $95\%$ thanks to micrometric accuracy of the emulsions, which allows to observe the displaced vertex of photon conversion. 

\section{SND@LHC and new physics}
\label{sec:examples}

Below, we illustrate the potential of SND@LHC to probe FIPs via decays and scatterings by estimating the sensitivity to several models. We consider two experimental setups of the detector: one that will operate during Run 3, and a possible upgrade that will work during Run 4 (see~\cite{TP}). Their parameters are summarized in Table~\ref{tab:parameters}.
\begin{table}[]
    \centering
    \begin{tabular}{|c|c|c|c|}
    \hline Setup & $\mathcal{L},\text{ fb}^{-1}$ &  $l^{\text{scatt}}_{\text{det}},\text{ cm}$   & $l^{\text{decay}}_{\text{det}},\text{ cm}$  \\ \hline
        Setup 1 & 150 &  30  & 50  \\ \hline
        Setup 2  & 3000 &  75   &  125 \\ \hline
    \end{tabular}
    \caption{Experimental setups of the SND@LHC detector used in this work. The parameters are the integrated luminosity $\mathcal{L}$, the detector's length available for scatterings $l^{\text{scatt}}_{\text{det}}$, the detector's length available for decays $l^{\text{decay}}_{\text{det}}$. See text for details.}
    \label{tab:parameters}
\end{table}

\subsection{Scattering}
\label{sec:scatterings}

In this subsection, we discuss LDM scattering off protons, and show that different signatures may be searched for, depending on the mediator mass. We apply our findings and estimate the expected sensitivity to the model of LDM interacting via leptophobic mediator.

\subsubsection{Elastic signature} 
\label{sec:elastic-inelastic}
There are two types of scattering off protons: elastic and inelastic, producing an isolated proton or hadronic showers, respectively. 

\begin{figure}[!h]
    \centering
    \includegraphics[width=0.5\textwidth]{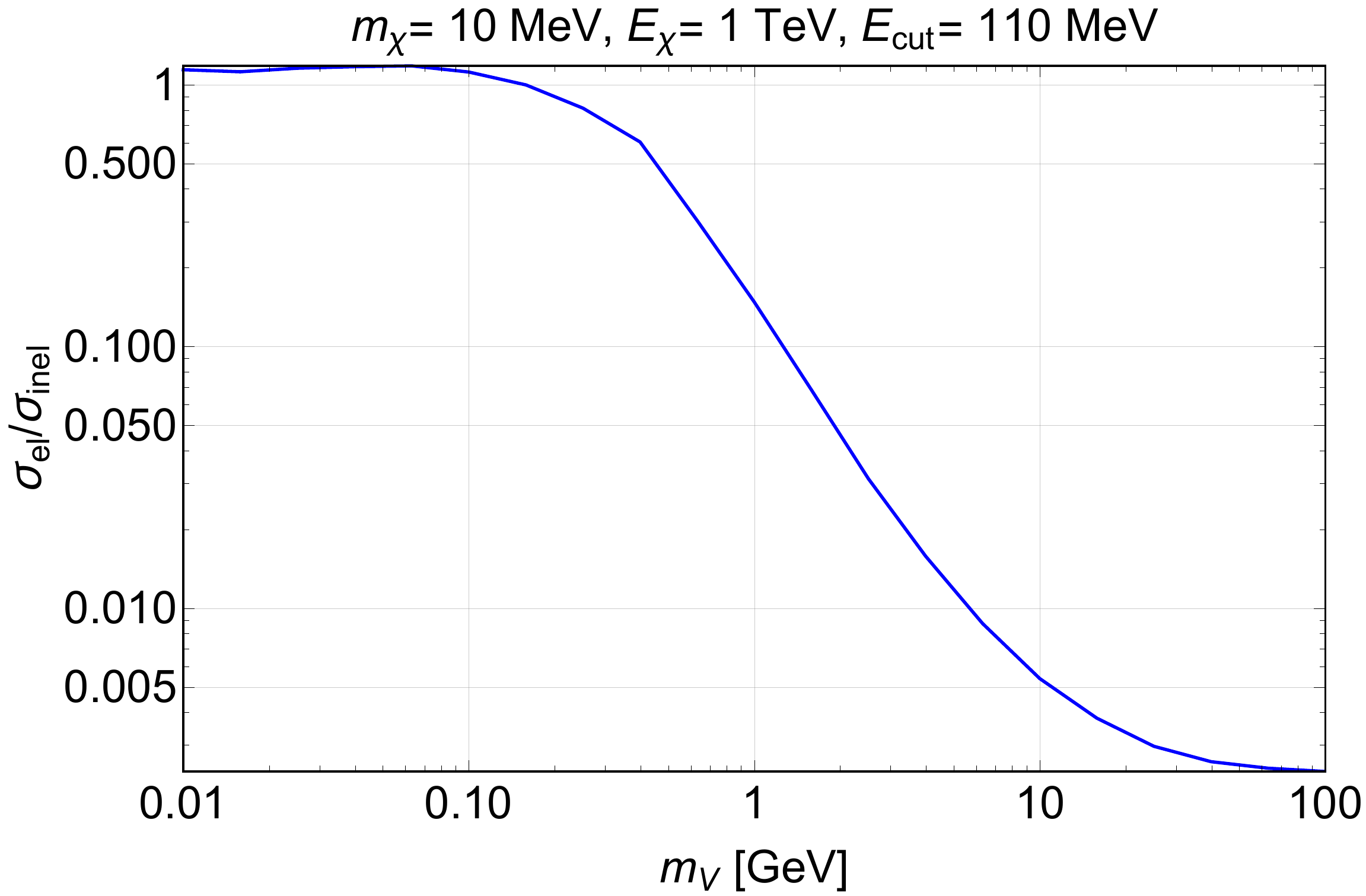}~\includegraphics[width=0.5\textwidth]{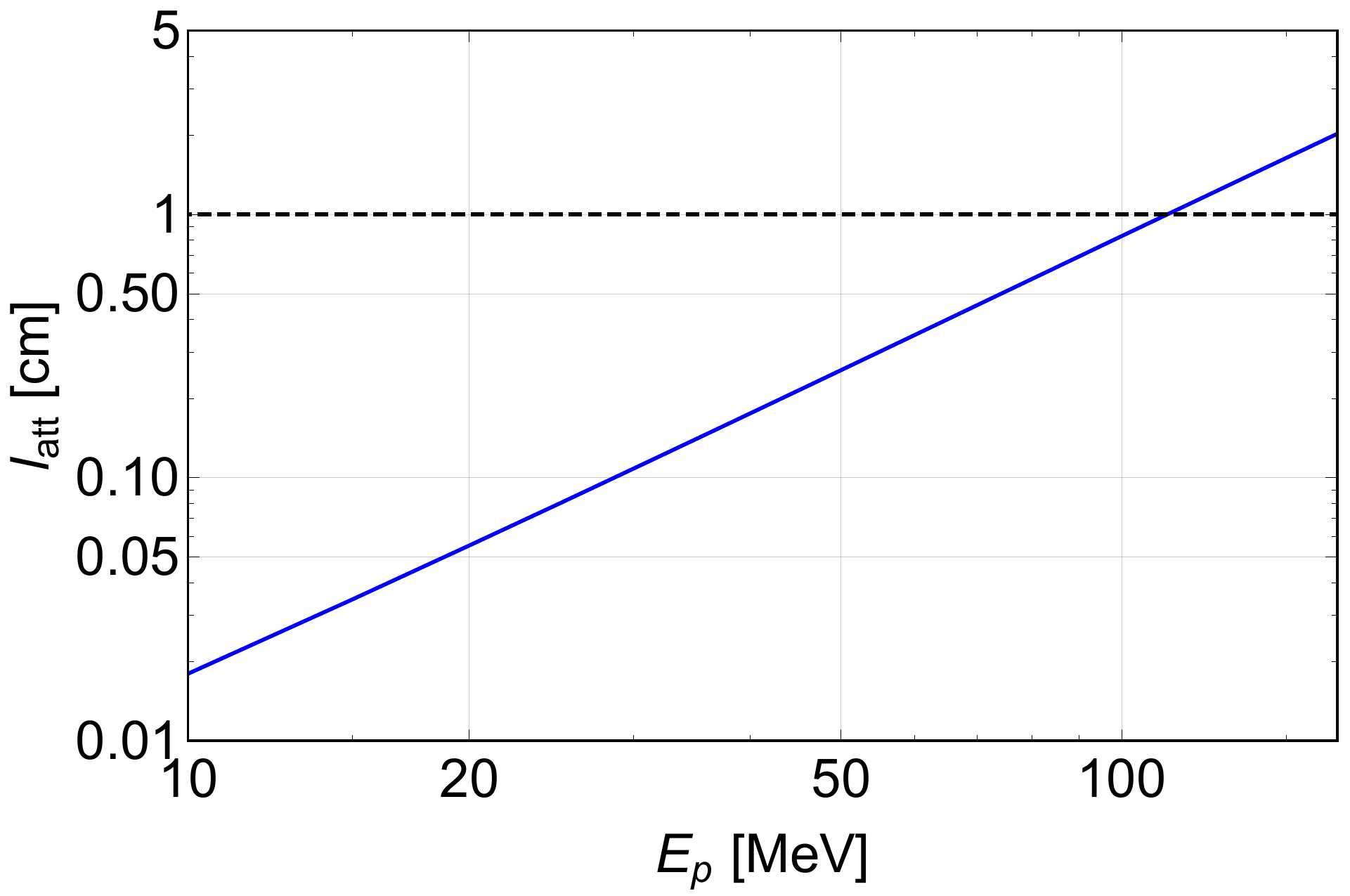}
    \caption{\textit{Left panel}: the ratio $\sigma_{\text{el}}/\sigma_{\text{inel}}$ of elastic and inelastic scattering cross sections in the model with a vector mediator $V$ interacting with protons and a scalar dark sector particle $\chi$ of mass $m_\chi = \unit[10]{MeV}$ and energy $E_\chi = \unit[1]{TeV}$. The minimal proton kinetic energy $E_{\text{cut}}\gtrsim \unit[110]{MeV}$ is assumed, for which protons may travel 1 cm in tungsten before being absorbed (see text for details). For the description of the elastic and deep inelastic scattering (DIS) used in the estimates, see Appendix~\ref{app:scattering-cross sections}. \textit{Right panel:} proton's attenuation length ($l_{\text{att}}=\int^{E_p}_0 \frac{dE}{dE/dx }$, where $dE/dx$ is the energy loss per unit length) in tungsten as a function of its kinetic energy. The value is calculated using the data from~\cite{StoppingPower}.}
    \label{fig:elastic-to-dis-ratio}
\end{figure}

In general, the ratio $N_{\text{el}}/N_{\text{inel}}$ for LDM interactions is different from that for NC neutrino interactions, decreasing with the growth of the mass of the mediator $m_{V}$. Indeed, both elastic and inelastic differential cross sections depend on the mediator mass $m_V$ as $d\sigma/d\Omega \propto (Q^2+m^2_V)^{-2}$ due to the propagator, where $Q^{2}$ is the momentum transfer. However, the elastic cross section also includes the proton form factor that limits the possible momentum transfer to $Q^2\lesssim r_{p}^{-2}\simeq \unit[1]{GeV^2}$. For large masses $m_{V}$, this leads to an additional suppression as compared to the inelastic cross section, to which all $Q^2\lesssim m_{V}^{2}$ contribute without the suppression~\cite{Dobrescu:2014ita}. As a result, the ratio $\sigma_{\text{el}}/\sigma_{\text{inel}}$ is a decreasing function of $m_{V}$. We illustrate this feature in Fig.~\ref{fig:elastic-to-dis-ratio}, considering a model of a scalar LDM that interacts with protons via a vector-like mediator. We see that in the case of light mediator $m_{V}\lesssim \unit[1]{GeV}$, the elastic and inelastic scattering yields may be comparable. However, with the increase of $m_{V}$, $\sigma_{\text{el}}/\sigma_{\text{inel}}$ quickly diminishes, and the inelastic events start to dominate. This is the case, for instance, for active neutrinos, for which the mediator is a $Z$ boson with $m_{Z}\approx 91\text{ GeV}$, and $\sigma_{\text{el}}/\sigma_{\text{inel}} \sim 10^{-3}$. This implies that the elastic signature (excess of elastic NC-like events) may have good sensitivity due to much smaller background from elastic neutrino scatterings.

Let us now discuss the signal reconstruction, efficiency, and background. 

The first question is whether the proton may be detected in emulsion and SciFi detector. It is important since protons from elastic scattering typically have low kinetic energy $E_{p}\ll 1\text{ GeV}$, since large energy transfer from LDM to the proton is suppressed by the form-factor and propagator. This means that protons cannot be reconstructed in SciFi detector alone, which may properly reconstruct high-energy particles only~\cite{TP}. Therefore, we restrict searches for the elastic signature to the emulsion. In order to be detected in the emulsion, the proton has to be energetic enough to fly through a few emulsion layers before being absorbed in the interleaving tungsten plates. We require the signal protons to have the kinetic energy $E_p \geq \unit[110]{MeV}$, for which the proton attenuation length in tungsten is larger than $\unit[1]{cm}$~\cite{Tanabashi:2018oca,RAYAPROLU201629}, see Fig.~\ref{fig:elastic-to-dis-ratio}, which corresponds to passing 10 emulsion layers by the proton before being absorbed. Even if passing 10 layers, the proton may be not reconstructed. To study this question, detailed Monte Carlo simulations are required. This goes beyond the scope of our paper, and we optimistically assume unity emulsion reconstruction efficiency.

The next question is background. MC simulations of neutrino scatterings from~\cite{TP} including the interaction of the reaction products with the detector have shown that during the operation time of SND@LHC the total number of neutrino NC resonant and deep inelastic scatterings where only one charged track is visible is $1.7$. However, what has not been studied is the background from the reaction products which cannot be properly associated to the neutrino or muon vertex and may therefore mimic the elastic signature signal. The whole emulsion is filled by tracks from muons and neutrino scatterings. Namely, the density of muon tracks in the bricks is about $\unit[5\cdot 10^4]{cm^{-2}}$~\cite{TP}. In addition, 12 neutrino DIS events are expected per each brick during half of the year; the hadronic showers in DIS events have transverse size of order of interaction length $\lambda_{I}\sim \unit[10]{cm}$, thus covering a large part of the volume of a brick. Some of the tracks from these cascades will not reach the electronic detectors because of being absorbed. Therefore, they are not associated to the vertex and may look similar to the elastic signature. The ability of the emulsion to associate tracks to the vertex, which may reduce such background to some extent, has not yet been studied.

The background may be suppressed by requiring isolation in some radius around the candidate proton track. The radius should be optimized taking into account occupancy and track reconstruction efficiency in the emulsion. Dedicated studies can be performed by the SND@LHC collaboration with full MC simulation and with first data expected to be collected in 2022. In principle, this problem could be mitigated if the emulsion bricks are replaced and used for data reconstruction during smaller time intervals, resulting in a lower number of background tracks.

\subsubsection{NC/CC signature}
\label{sec:charged-neutral}
For masses $m_V \gtrsim m_p$, LDM is more likely to scatter inelastically. In this case, we need to see these events over the numerous neutrino scattering background. However, the total number of NC neutrino events is a subject of theoretical uncertainties. It depends on the yield of neutrinos flying in the direction of SND@LHC, which is determined by the production of mesons in the far-forward direction.\footnote{One of goals of SND@LHC is to study the production of mesons in the far-forward direction, which is currently poorly probed via experiments.} At the same time, the ratio of neutral current and charged current events $N_\text{NC}/N_\text{CC}$ for neutrinos is uniquely predicted within the SM. For the tungsten target, under the approximation of equal differential distributions of $\nu$ and $\bar\nu$, the ratio $N_\text{NC}/N_\text{CC}$ for deep inelastic scattering is equal to~\cite{TP}
\begin{equation}
    P = \frac{1}{2}\left[1-2\sin^2 \theta_W+\frac{20}{9}\sin^4\theta_W - \lambda (1-2\sin^2\theta_W)\sin^2 \theta_W\right] \approx 0.33
\end{equation}
where $\lambda = 0.040$ for the tungsten target. The overall uncertainty in the $P$ value measurement at the SND@LHC is expected to be $10\%$. Assuming $N_{\text{CC}}^{\text{SND@LHC}} = 1395$ and $N_{\text{NC}} = 450$ as predicted by simulations for the SND@LHC setup~\cite{TP}, we require the yield of LDM inelastic scattering events to be $2\sqrt{450 + 45.0^2} \approx 100$ in order to reach the 2$\sigma$ confidence level.

\begin{figure}
    \centering
    \includegraphics[width = 0.7\textwidth]{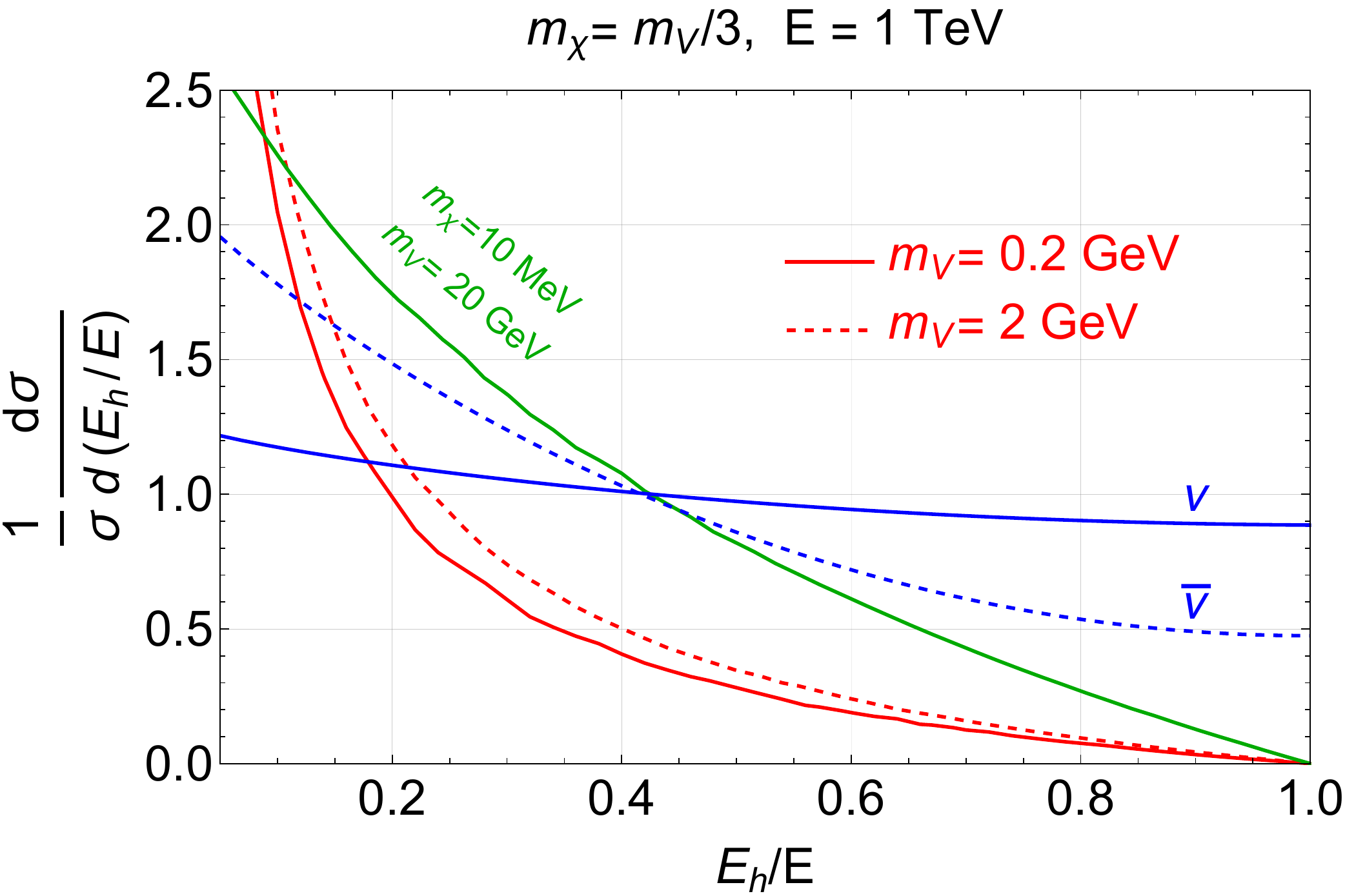}
    \caption{Spectra of $\nu$, $\bar\nu$, (NC) and $\chi$ inelastic scattering events. Here, $E$ is the energy of an incoming particle, and $E_h$ is the total energy transferred to produced hadrons. The form of these spectra does not change significantly for different values of $E$ in the energy range of interest. The green line shows the special case of light $\chi$ and massive $V$.}
    \label{fig:DISspectra}
\end{figure}

LDM scattering may also be distinguished from neutrino events due to different kinematics and topology of the final state hadrons. Therefore, energy requirements may additionally reduce the neutrino background, see Fig.~\ref{fig:DISspectra}. The detailed analysis of this question is the subject of a separate study.

\subsubsection{Example: leptophobic DM}
\label{sec:leptophobic}
Let us now estimate the sensitivity of SND@LHC to the elastic/inelastic and NC/CC signatures for a particular model of LDM. We consider a theory with a scalar particle~$\chi$ coupled to the SM via a vector mediator $V$ that interacts with the baryon current $J_{\mu}^{B}$~\cite{Batell:2014yra,Dobrescu:2014ita,deNiverville:2016rqh,Buonocore:2018xjk,Coloma:2015pih}:
\begin{equation}
    \mathcal{L}_{\text{leptophob}} = -g_{B}V^{\mu}J_{\mu}^{B} + g_{\chi}V^{\mu}(\partial_{\mu}\chi^{\dagger}\chi-\chi^{\dagger}\partial_{\mu}\chi), \quad J^{B}_{\mu} = \frac{1}{3}\sum_{q}\bar{q}\gamma_{\mu}q
    \label{eq:leptophobic-portal}
\end{equation}
Here, $g_{\chi},g_{B}$ are coupling constants of the mediator to $\chi$ and SM sector, and the sum in $J^{\mu}_{\mu}$ is made over all quark flavors.

\begin{figure}[!h]
    \centering
    \includegraphics[width=0.5\textwidth]{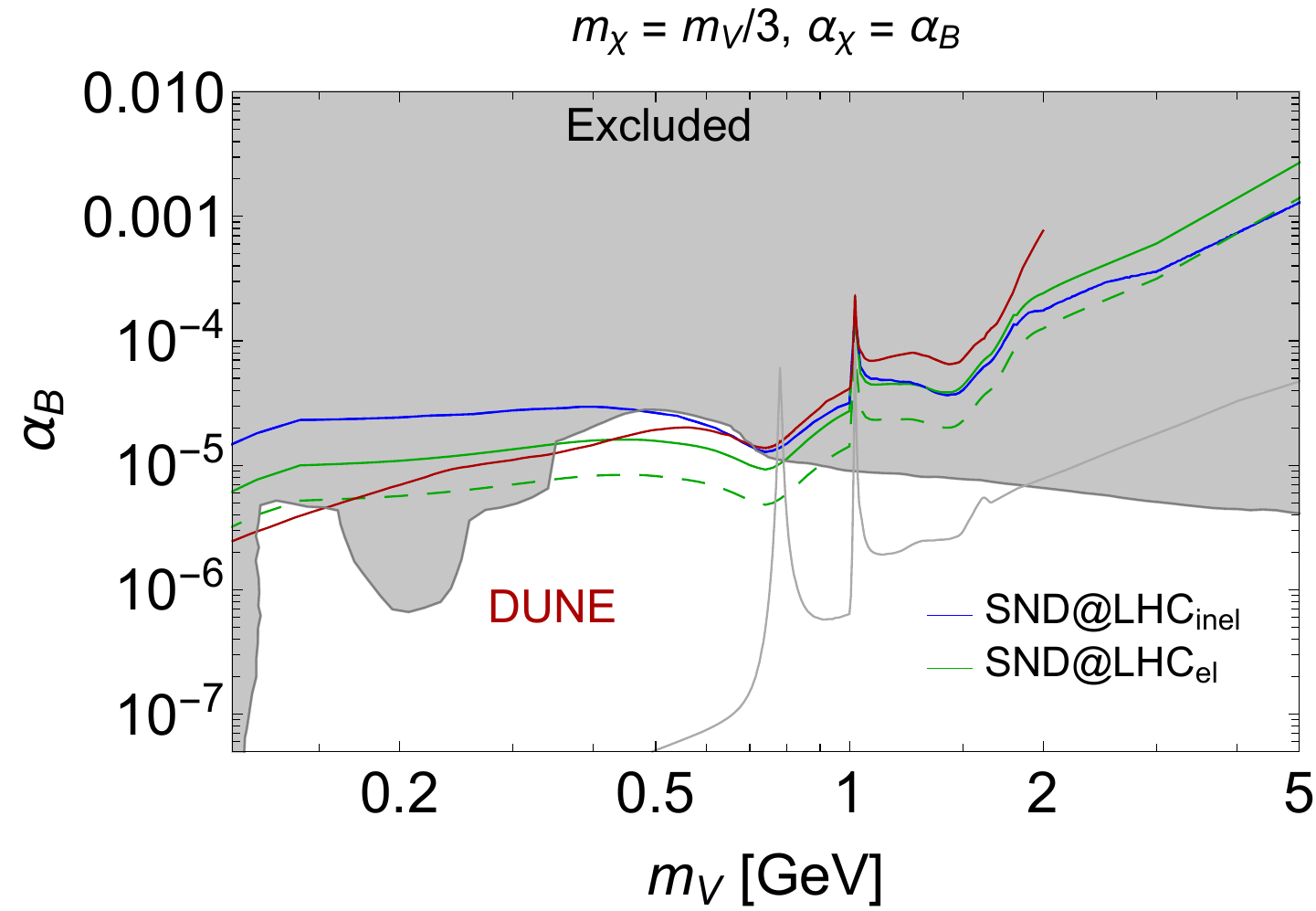}~\includegraphics[width=0.5\textwidth]{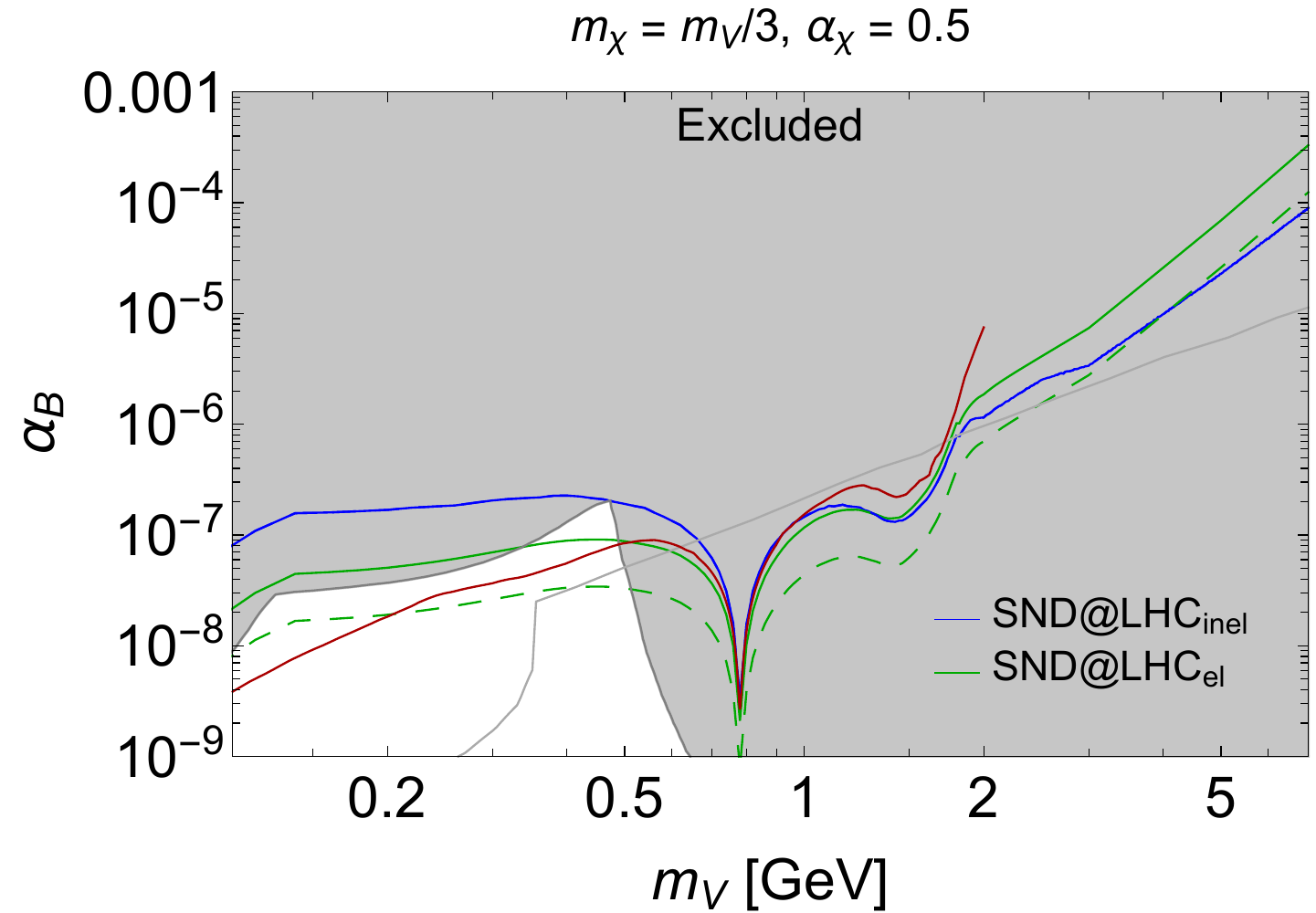}
    \vspace{1mm}
    
    \includegraphics[width=0.5\textwidth]{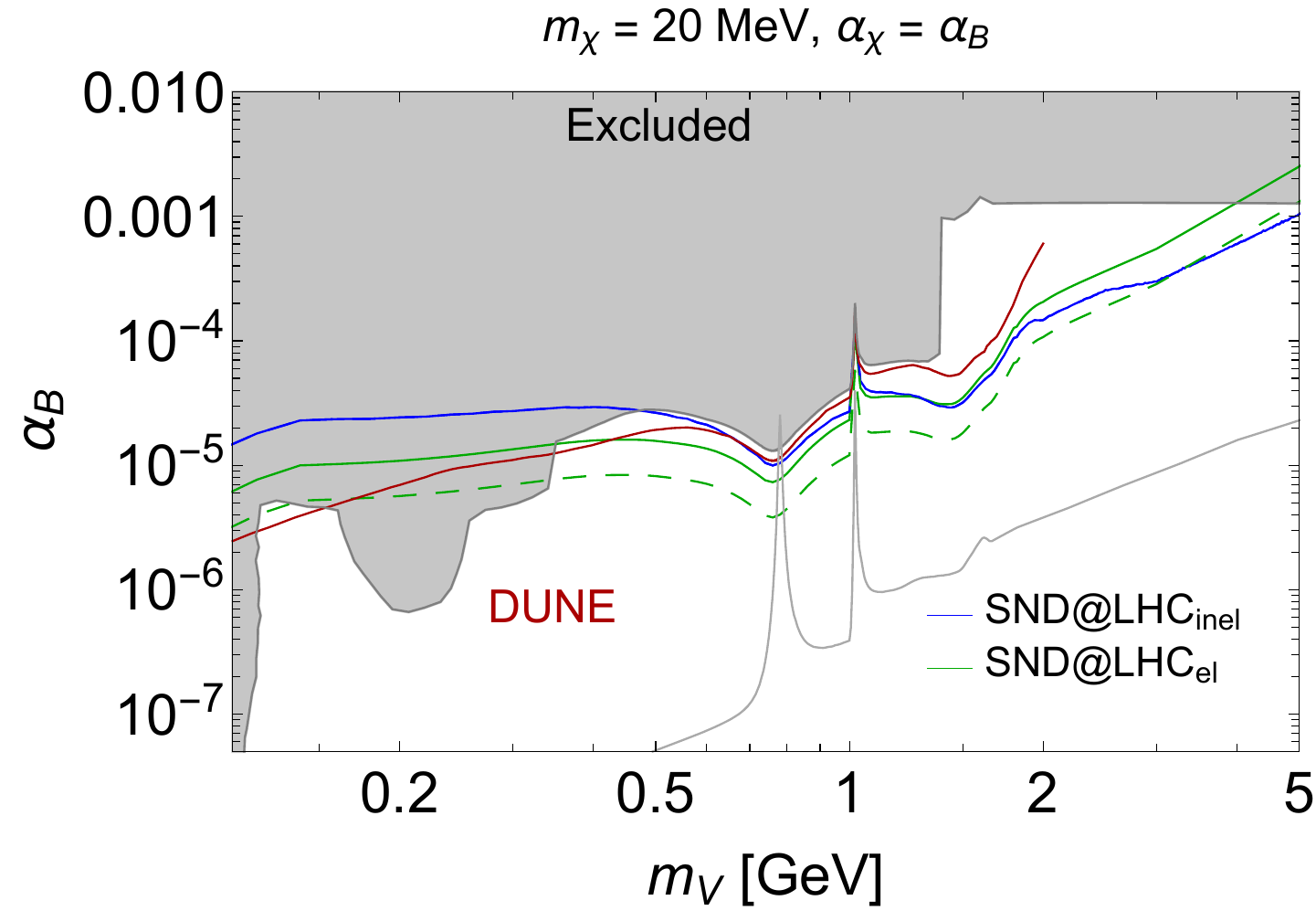}~\includegraphics[width=0.5\textwidth]{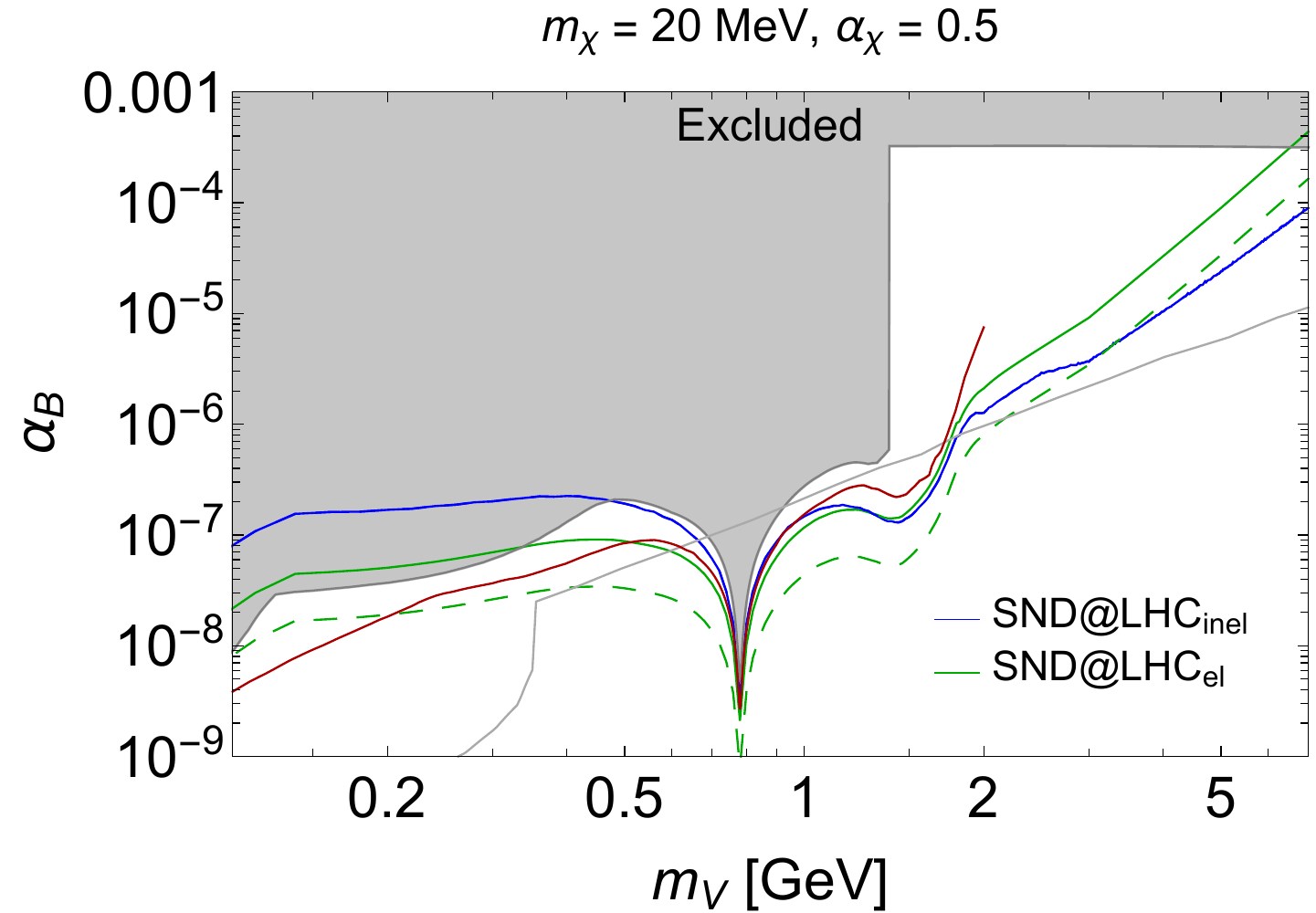}
    \caption{Sensitivity of the SND@LHC experiment to the leptophobic portal~\eqref{eq:leptophobic-portal} ($2\sigma$ CL). The sensitivity is shown under an assumption $m_{\chi} = m_{V}/3$ (top panel) and $m_{\chi} = \unit[20]{MeV}$ (bottom panel), and for two different choices of the coupling of mediator to $\chi$ particles: $\alpha_{\chi} = \alpha_{B}$ (left figures), and $\alpha_{\chi} = 0.5$ (right figures). The considered signatures are the elastic scattering off protons (the green line) and the deep-inelastic scattering (the blue line, corresponding to 100 signal events during Run 3). For the elastic signature we show contours for 10 events; the solid line corresponds to the setup during Run 3, while the dashed line corresponds to the upgraded setup that may operate during Run 4. We stress that the result for the elastic signature has to be taken with caution, since it crucially depends on background and signal identification for the elastic signature, which has not been performed yet for SND@LHC and may significantly reduce the elastic sensitivity (see Sec.~\ref{sec:elastic-inelastic} for details). 
    By the red line, we show the 100 event contour for the DUNE experiment from Ref.~\cite{Naaz:2020tgx}. We rescale the previous bounds according to our description of the proton form-factor used in bremsstrahlung and $\text{Br}(V\to \chi\chi)$. The thin gray line corresponds to model-dependent constraints from invisible decays~\eqref{eq:FCNC} as derived in~\cite{Ilten:2018crw} (see text for details).}
    \label{fig:leptophobic-sensitivity}
\end{figure}

\textbf{Current constraints.} Constraints on the model~\eqref{eq:leptophobic-portal} are summarized in Fig.~\ref{fig:leptophobic-sensitivity}. For $m_V\gtrsim \unit[0.1]{GeV}$, they come from searches for decays $\pi,K,\eta\to V\gamma$ at CB~\cite{Amsler:1994gt}, E949~\cite{Artamonov:2009sz}, and NA62~\cite{CortinaGil:2019nuo} experiments (for $m_{V}\lesssim \unit[0.5]{GeV}$), searches for scattering of $\chi$ particles off nucleons at MiniBooNE~\cite{Aguilar-Arevalo:2018wea} (for $\unit[0.5]{GeV}\lesssim m_{V}\lesssim \unit[1.5]{GeV}$), a monojet signature analysis at CDF~\cite{Aaltonen:2012jb} (for $m_{V} \gtrsim\unit[1.5]{GeV}$), and direct DM searches at CRESST III~\cite{Abdelhameed:2019hmk}. 

The weakness of the CDF monojet signature and the absence of direct constraints from the LHC is caused by the requirement of a large missing transverse momentum $p_{T}\sim \unit[100]{GeV}$ for signal tagging and background suppression. Such large $p_{T}$ may be provided only by large mass of a decaying particles, which is definitely not the case of light $\mathcal{O}(\unit[1]{GeV})$ mediators considered in this paper. The bounds from MiniBooNE, being one of the strongest in the region $m_{V}\lesssim \unit[1]{GeV}$, are much weaker at larger masses due to small center-of-mass energy of the $pp$ collisions, $\sqrt{s}\approx\unit[4]{GeV}$.

Another constraint comes from DM direct detection experiments (DD)~\cite{Batell:2014yra} that search for scattering of DM particles off nuclei. The sensitivity of these experiments depends on DM particle mass. Indeed, it determines the maximal kinetic energy of DM (which is $T_{\chi} = m_{\chi}v_{\text{escape}}^{2}/2$, where $v_{\text{escape}} = \unit[544]{km/s}$ is the escape velocity), and, therefore, the maximally possible nuclear recoil energy $T_{N}$. The DD experiments have finite energy threshold, being $T_{N}>\unit[30.1]{eV}$ for CRESST-III~\cite{Abdelhameed:2019hmk} that is currently the most sensitive experiment. As a result, current constraints from DD are limited by $m_\chi \gtrsim \unit[160]{MeV}$. In addition, the DD bounds may be significantly relaxed even for heavy $\chi$ particles if assume their axial-vector interaction with $V$ instead of vector-like one (see~\cite{Caron:2018yzp}) that results in the velocity-suppressed scattering cross section.

Finally, in~\cite{Dror:2017nsg,Dror:2017ehi,Ilten:2018crw}, it was argued that the strongest constraint may come from negative results of searches for decays
\begin{equation}
K \to \pi + \text{inv}, \ B\to K+\text{inv}, \ Z\to \gamma+\text{inv}
\label{eq:FCNC}
\end{equation}
at LHCb. In the model of the leptophobic portal~\eqref{eq:leptophobic-portal}, the decays~\eqref{eq:FCNC} may result from the anomalous violation of the baryon current conservation, which requires a UV completion in order to cancel the anomaly. Namely, in~\cite{Dror:2017nsg,Dror:2017ehi}, it was considered a UV completion with some heavy fermions such that the full theory is anomaly-free. At energies much lower than masses of these fermions, the effective theory contains, apart from the Lagrangian~\eqref{eq:leptophobic-portal}, pseudo-Chern-Simons (pCS) interaction operators between $V$ and electroweak bosons $W,Z,\gamma$ that result from the contribution of massive fermions to the anomalous triangle diagrams. The latter contain two summands: a mass-independent, and a mass-dependent. The sum of the first terms over all fermions vanishes due to the anomaly cancellation, while the net mass-dependent part is in general non-zero (for instance, if there is a hierarchy in fermion masses). The corresponding interactions mediate the process $Z\to \gamma +X$, and generate effective flavor changing neutral current couplings $bsV$, $sdV$ between quarks and the leptophobic mediator (via penguin loop diagrams) that mediate the first two processes in Eq.~\eqref{eq:FCNC}.

pCS terms generically appear in effective theories with chiral fermions. However, their contribution to the processes~\eqref{eq:FCNC} depends on the UV completion of the model~\eqref{eq:leptophobic-portal}. For instance, one could consider a 3+n+1 dimensional model with SM physics localized on a 3+1 dimensional sub-manifold (brane) and a large mass gap for the bulk modes (see e.g.~\cite{Boyarsky:2005hs,Boyarsky:2005eq}). The higher-dimensional theory is anomaly free by construction without adding extra fermions. The anomaly of the low-dimensional 3+1 effective theory is done by the  ``anomaly inflow'' mechanism, non-local from 3+1 dimensional point of view. In this case, the anomaly cancellation by massive modes does not contribute to decays.

Due to the model dependence, the status of the anomaly constraint is different from the status of the other bounds discussed above, as the latter require only the effective Lagrangian~\eqref{eq:leptophobic-portal}.  Therefore, in Fig.~\ref{fig:leptophobic-sensitivity}, where we show the constraints on the leptophobic portal, we just indicate the parameter space potentially constrained by processes~\eqref{eq:FCNC} by showing its lower bound only, while for the other constraints discussed in this subsection the whole parameter space is shown in solid gray.

\vspace{5mm}
\textbf{Number of events.} Let us now estimate the sensitivity of SND@LHC to LDM scattering in the model~\eqref{eq:leptophobic-portal}. The number of scattering events may be estimated using the formula
\begin{equation}
    N_{\text{events}} = 2\cdot N_{\chi}^{\text{SND@LHC}}\times n_{\text{detector}}\times \begin{cases}Z\cdot \sigma_{\text{scatt}}^{\text{el}}(\langle E_{\chi}\rangle)\cdot l^\text{scatt}_\text{det} ,\quad \text{elastic signature}\\ A\cdot \sigma_{\text{scatt}}^{\text{inel}}(\langle E_{\chi}\rangle)\cdot l^\text{scatt}_\text{det}, \quad \text{NC/CC signature} \end{cases}
    \label{eq:number-of-events-scattering}
\end{equation}
Here, $N_{\chi}^{\text{SND@LHC}}$ is the number of $\chi$ particles produced in the direction of the SND@LHC detector volume (a factor of 2 stays for $\bar{\chi}$), $n_{\text{detector}}$ is the detector's atomic number density (the tungsten material is considered), $Z,A$ are atomic and mass numbers of the target material, and $\sigma^{\text{el/inel}}_{\text{scatt}}$ is the elastic or inelastic scattering cross section of $\chi$ particles. For simplicity, when calculating the cross section, we assume that all $\chi$ particles have the same energy equal to their average energy $\langle E_{\chi}\rangle$.

We adopt the description of the elastic scattering process from~\cite{deNiverville:2016rqh}. We do not take into account further re-scattering of the produced proton, which potentially may lead to appearance of charged tracks and affect the signal selection. For the estimate of the cross-section for inelastic scattering, we use the calculation based on the parton model from~\cite{Soper:2014ska}, for which parton distribution functions are given by CT10nlo PDF sets from LHAPDF package~\cite{Buckley:2014ana} (see also Appendix~\ref{app:scattering-cross sections}).

\begin{figure}[!h]
    \centering
    \includegraphics[width=0.9\textwidth]{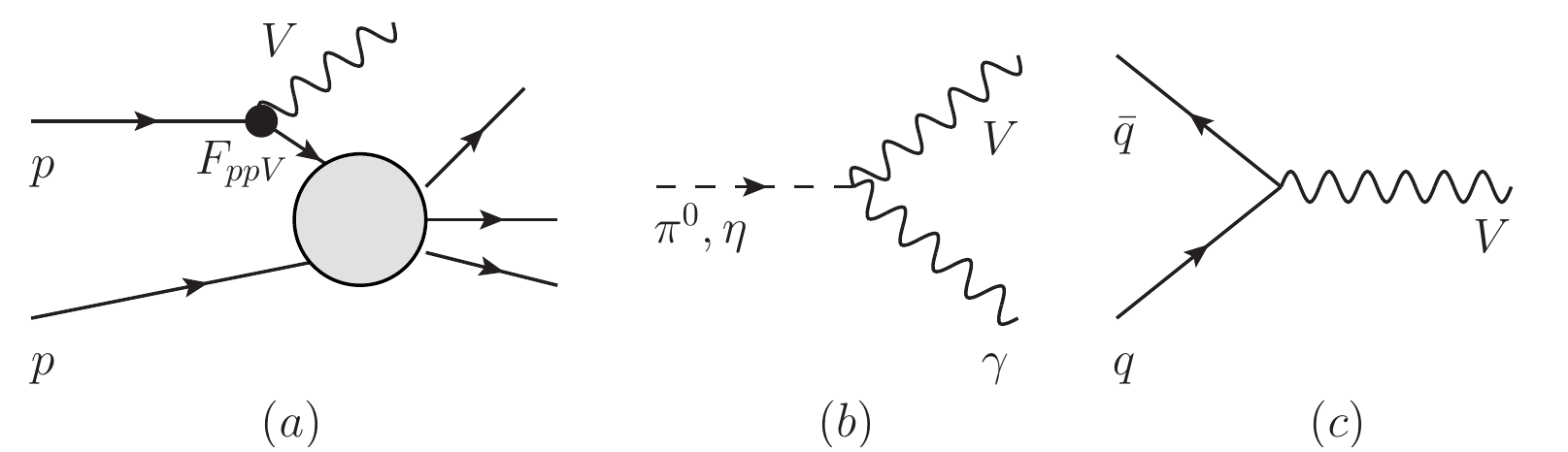}
    \caption{Diagrams of the production of the leptophobic mediator $V$: by proton bremsstrahlung  (a), in decays of light unflavored mesons (b), and in Drell-Yan process (c).}
    \label{fig:leptophobic-mediator-production}
\end{figure}

Let us now consider the production of $\chi$ particles. The $\chi\bar{\chi}$ pairs originate from decays of $V$. Similarly to the dark photon case, the mediator may be produced:
\begin{enumerate}
\item in decays of unflavored mesons $\pi,\eta$,
\begin{equation}
    \pi \to V + \gamma, \quad \eta \to V + \gamma,
\end{equation}
\item by proton bremsstrahlung, 
\begin{equation}
p + p \to V+X,
\end{equation}
\item in Drell-Yan process, \begin{equation}
q+\bar{q} \to V+X,
\end{equation}
\end{enumerate}
see Fig.~\ref{fig:leptophobic-mediator-production}.
For the description of these channels, we mainly follow~\cite{Tulin:2014tya,Ilten:2018crw,Kling:2020iar}. 

For the production from mesons, we use the polar angle and energy distributions of $\pi$, $\eta$ mesons generated by EPOS-LHC~\cite{Pierog:2013ria} as a part of the CRMC package~\cite{CRMC}. The resulting spectra of $V$ and $\chi$ particles are obtained semi-analytically using an approach presented in~\cite{Boiarska:2019vid}. 
\begin{figure}[!h]
    \centering
    \includegraphics[width=0.5\textwidth]{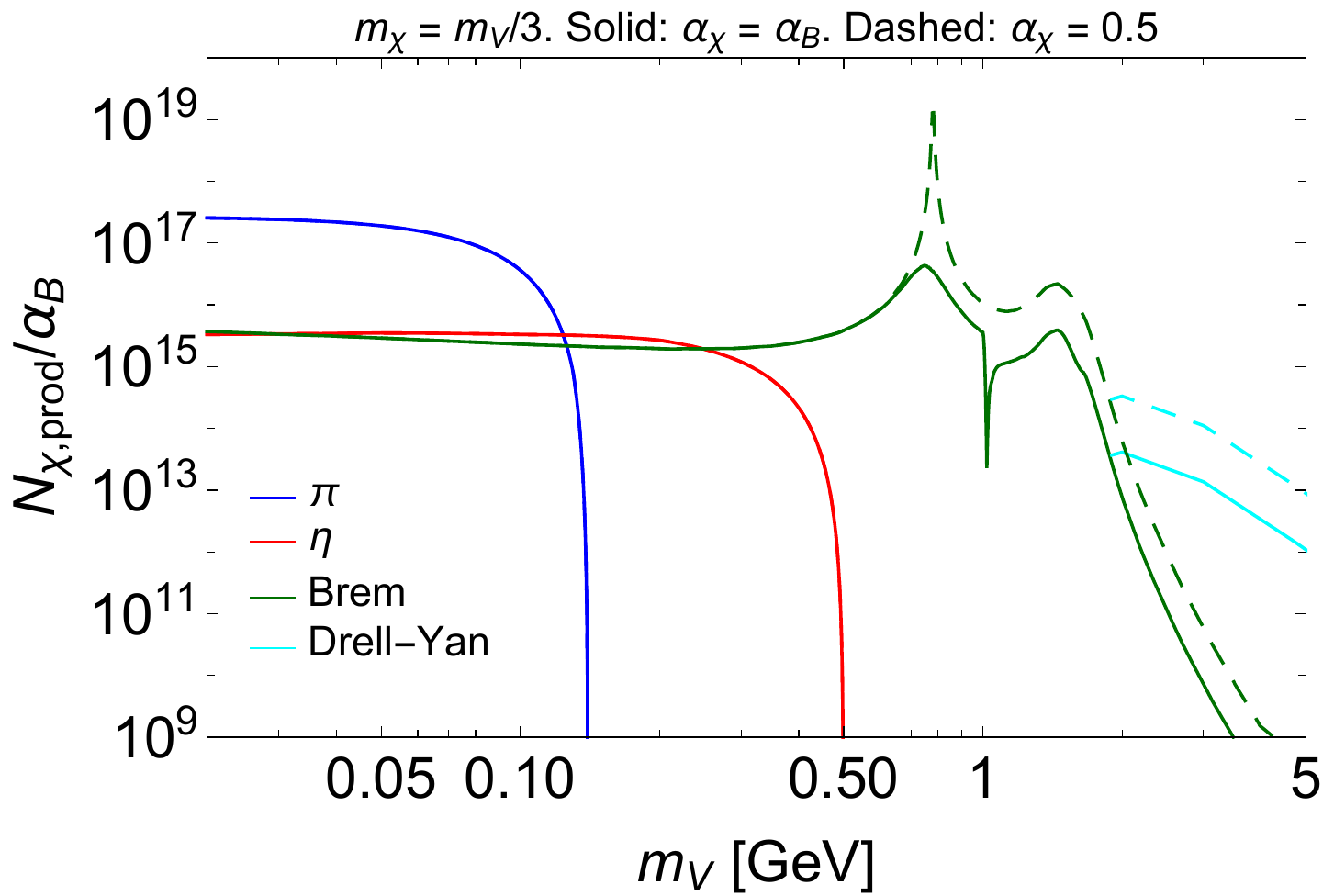}
    \caption{The number of $\chi$ particles produced in the direction of the SND@LHC experiment, assuming the integrated luminosity $\mathcal{L} = \unit[150]{ fb^{-1}}$. $m_{\chi} = m_{V}/3$ is assumed. Wiggles around $V$ masses of $\unit[782]{MeV}, \unit[1020]{MeV}$ and $\simeq \unit[1.7]{GeV}$ are caused by the mixing of the mediator with isoscalar vector mesons $\omega$, $\phi$, and their excitations, which leads to the resonant enhancement of 1) the $ppV$ form-factor for the production by the proton bremsstrahlung, and 2) the decay width of the leptophobic mediator $V$ into hadrons (and hence to a suppression of $\text{Br}(V\to \chi\bar{\chi})$). See text and Appendix~\ref{app:leptophobic-phenomenology} for details.}
    \label{fig:V-prod}
\end{figure}

For obtaining the angle-energy distribution of the leptophobic mediator produced by the proton bremsstrahlung, we consider the kinematic range $p_{T}<\unit[1]{GeV}$ and $0.1<z<0.9$. The corresponding production probability is affected by the mixing of $V$ with isoscalar $\omega$ and $\phi$ mesons. To describe this effect, we follow the procedure described in~\cite{Faessler:2009tn} (see also~\cite{Kling:2018wct} and Appendix~\ref{app:leptophobic-phenomenology} for details). The distribution of subsequent $\chi$ particles produced by the bremsstrahlung is obtained in a similar way as for the case of the production from mesons. 

For the production in the Drell-Yan process, we use our implementation of the model~\eqref{eq:leptophobic-portal} in MadGraph5~\cite{Alwall:2014hca} with FeynRules~\cite{Alloul:2013bka,Christensen:2008py}. We then obtain the geometric acceptance and energy distribution of $\chi$ particles traveling into the direction of the SND@LHC detector by simulating the leading-order process 
\mbox{$p + p \to V,\  V \to \chi \bar{\chi}$}. 

We find that the main production channel for masses $m_{V}\lesssim m_{\eta}$ is decays of mesons, for masses $m_{\eta}\lesssim m_{V}\lesssim \unit[3]{GeV}$ is the proton bremsstrahlung, and, finally, for $m_{V}\gtrsim \unit[3]{GeV}$ it is the Drell-Yan process, see Fig.~\ref{fig:V-prod}. 

Most of the produced $\chi$~particles have $\gamma$ factors $\sim 10^{3}$, independently of the production channel. This means that the time-of-flight measurement is not efficient in separating signal $\chi$ particles and neutrinos.
\vspace{5mm}

\textbf{Sensitivity.} Let us now discuss the sensitivity. The parameters in the model are LDM particle and mediator masses $m_{\chi,V}$, and the couplings $\alpha_{B} = g_{B}^{2}/4\pi, \alpha_{\chi} = g_{\chi}^{2}/4\pi$.

The choice of $\alpha_{\chi}$ affects the parameter space probed by SND@LHC in the following way. The number of scattering events at SND@LHC scales as 
\begin{equation}
N_{\text{events}}\propto \alpha_{B}\cdot \text{Br}(V\to \chi\bar{\chi}) \times \alpha_{B}\cdot \alpha_{\chi}
\end{equation}
Here, a factor $\alpha_{B}\cdot \text{Br}(V\to \chi\bar{\chi})$ comes from the production, while a factor $\alpha_{B}\cdot \alpha_{\chi}$ -- from the subsequent scattering of $\chi$ particles. The scaling of the previous bounds is somewhat different. While the scaling of events at MiniBooNE is similar, the number of events at the other experiments scales as $N_{\text{events}}\propto \alpha_{B}\cdot \text{Br}(V\to \chi\bar{\chi})$ for the collider experiments and $\alpha_{B}\cdot \alpha_{\chi}$ for DD experiments. Therefore, the dependence on $\alpha_{B}$ and $\alpha_{\chi}$ is weaker.\footnote{For the calculation of the branching ratio $\text{Br}(V\to \chi\bar{\chi})$, see Appendix~\ref{app:leptophobic-phenomenology}.} Therefore, marginalizing over $\alpha_{\chi}$, the optimal choice would be $\alpha_{\chi} \simeq 1$, for which SND@LHC would probe larger range of mediator masses. 

We consider two values of $\alpha_{\chi}$. The first one is $\alpha_{\chi} = \alpha_{B}$, which is typically considered in the literature, and for which $N_{\text{events}}\propto \alpha_{B}^{3}\cdot \text{Br}(V\to \chi\bar{\chi})$. The second one is $\alpha_{\chi} = 0.5$, for which $N_{\text{events}} \propto \alpha_{B}^{2}$. 

Let us now comment on the choice of $m_{\chi}$ for representing the results. As we have discussed previously, masses $m_{\chi}>\unit[160]{MeV}$ are significantly constrained by the DM direct detection experiments. Therefore, we consider two different choices: $m_{\chi} = m_{V}/3$, which is commonly used in literature and for which the DD constraint is important above $m_{V}=\unit[480]{MeV}$, and $m_{\chi} = \unit[20]{MeV}$, for which there is no bound from DD at all. The sensitivity of SND@LHC depends only weakly on this choice, as the production probability and the scattering cross section of high-energy $\chi$ particles is determined mainly by $m_{V}$. In this way, SND@LHC and direct DM detection experiments may probe complementary mass ranges of $\chi$.

The sensitivity of the SND@LHC experiment to the leptophobic portal for two different setups from Table~\ref{tab:parameters} is shown in Fig.~\ref{fig:leptophobic-sensitivity}. Following the discussion in Sec.~\ref{sec:charged-neutral}, we require $N_{\text{events}}>100$ for the NC/CC signature during Run 3. For the elastic scattering, we show the contours corresponding to $N_{\text{events}}>10$.

The parameter space that may be probed by SND varies in dependence on the values of parameters $\alpha_{\chi}$, $m_{\chi}$. Namely, for the choice $m_{\chi} = m_{V}/3$, using the elastic signature, SND@LHC only may probe masses $\unit[350]{MeV} < m_{V}< \unit[700]{MeV}$ if the experimental challenges with signal identification discussed in Sec.~\ref{sec:elastic-inelastic} may be overcome. For the choice $m_{\chi} = \unit[20]{MeV}$, using the NC/CC signature only, it is possible to probe masses $\unit[700]{MeV} < m_V \lesssim \unit[7]{GeV}$. Moreover, for the choice $\alpha_{\chi} = 0.5$, the probed range of the coupling $\alpha_{B}$ even competes with the model-dependent bound from the signature $B\to K+\text{inv}$ at the lower bound.

In the figure, we also show the sensitivity of DUNE experiment from~\cite{Naaz:2020tgx}. The background estimate has not been made for this experiment. Therefore, we show the contour corresponding to $100$ events.

\subsection{Decays}
\label{sec:decays}
As discussed in Sec.~\ref{sec:introduction},  the decay signature is decays of FIPs into a pair of charged particles. However, to use this signature, it is necessary to disentangle their tracks. At the SND@LHC detector, this is possible if the transverse distance between the tracks exceeds the spatial resolution, which is of the order of $\unit[1]{\mu m}$ for the emulsion films. The transverse distance between two tracks is determined by the flight angle that can be estimated as $\Delta \theta \simeq \frac{m_{\text{FIP}}}{E_{\text{FIP}}}$, and the distance $l$ charged particles travel inside the target. For electrons, $l$ is the radiation length, which in tungsten is equal to $\unit[3.5]{mm}$. Muons pass through the whole target without deflection, and therefore, we may conservatively restrict $l$ to the thickness of a single SND@LHC emulsion brick $\unit[7.8]{cm}$. For FIPs flying in the far-forward direction, the typical energy is $E_{\text{FIP}}\simeq \unit[1]{TeV}$. Thus, the corresponding masses are 
\begin{equation}
m_{\text{FIP}} \gtrsim \text{max}\left[E_{\text{FIP}}\, \frac{\unit[1]{\mu m}}{l}, 2m_{e\text{ or }\mu}\right] \simeq \begin{cases}
\unit[290]{MeV}, & \text{FIP}\to e\bar e\\
\unit[210]{MeV}, & \text{FIP}\to \mu\bar \mu
\end{cases}
\end{equation}
If the disentanglement is not possible, instead of tracks we observe a mono-cascade. A similar signature may come from FIPs decaying into neutral pions, such as from HNLs that mix with tau flavor that decay into $\pi^{0}$ and a neutrino. This type of events may still be distinguished from neutrino scatterings, as the latter typically contain many tracks, and hence may be a new physics signature.

Background evaluation and event reconstruction for both charged pair and monocascade signatures are challenging tasks and require dedicated studies. Further, we will show the fixed signal events contours, assuming that all events are detected.

\subsubsection{Sensitivity to portals}
\label{sec:portals}
To illustrate the potential of SND@LHC to probe decays, we estimate the sensitivity to scalar, neutrino and vector portals, which introduce correspondingly a light Higgs-like scalar, a heavy neutral lepton (HNL) and a dark photon (see, e.g.,~\cite{Beacham:2019nyx} for the description of the models). Decays with pairs of charged particles in the final state -- muons, electrons and pions -- are the main decay channels for all the portal particles, except for GeV scale HNLs $N$ that mix with $\nu_{\tau}$, for which the main decay channel is a decay $N\to \pi^{0}\nu$.

In order to obtain the sensitivity of SND@LHC to various decaying FIPs, we use the following estimate:
\begin{equation}
    \nev = \sum_{\text{i}}N^{i}_{\text{prod}} \cdot \egeom^{i}\cdot \pdec^{i} \cdot \brvis
\end{equation}

\begin{figure}[!h]
    \centering
    \includegraphics[width=0.45\textwidth]{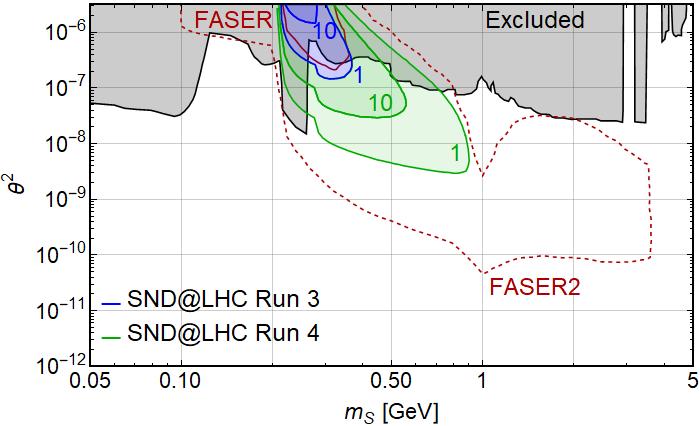}~\includegraphics[width=0.45\textwidth]{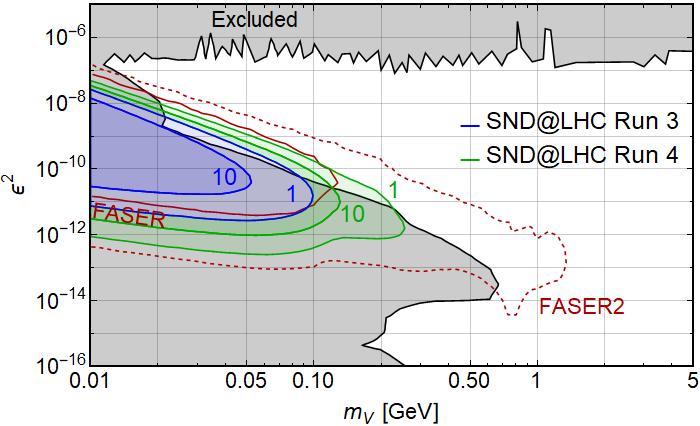}\\\includegraphics[width=0.45\textwidth]{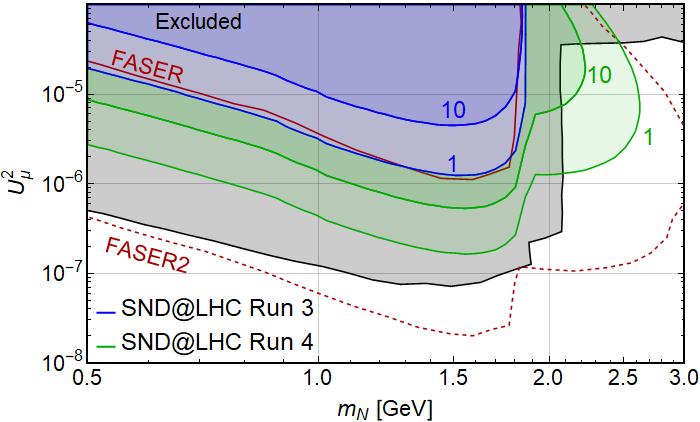}~\includegraphics[width=0.45\textwidth]{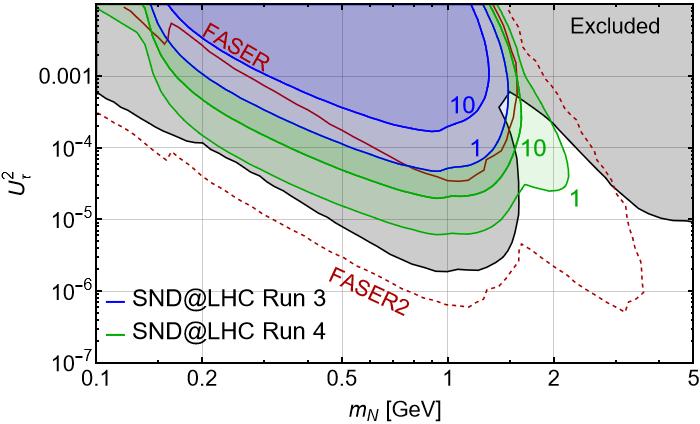}
    \caption{Signal rate contours of SND@LHC to (top left) dark scalars, (top right) dark photons, and HNLs that mix with (bottom left) $\nu_{\mu}$ and (bottom right) $\nu_{\tau}$. Blue (green) contours correspond to 1 and 10 events in the (upgraded) SND@LHC target. The actual sensitivity of SND@LHC can be derived taking into account the signal identification efficiency and background level, which are subjects of detailed studies that go beyond the scope of this paper, see Sec.~\ref{sec:decays}. Sensitivities of previous experiments and of the FASER/FASER2 experiment are reproduced from~\cite{Beacham:2019nyx,Boiarska:2021yho}.}
    \label{fig:sensitivity-to-decays}
\end{figure}

Here, $N_{\text{prod}}^{i}$ is the total number of FIPs of species $X$ produced via channel $i$, $\egeom^{i}$ is the geometric acceptance for particle $X$ decay products, and $\pdec^{i}$ is the decay probability averaged over energies $E_{X}$ of particles $X$, 
\begin{equation}
    \pdec^{i} = \int (e^{-\lmin/c\tau_{X}\gamma_{X}} - e^{-\lmax/c\tau_{X}\gamma_{X}})f^{i}_{E_{X}}\text{d}E_{X},
\end{equation}
with $f_{E_{X}}$ being the energy $E_X$ distribution of FIPs that fly in the decay volume, and $\tau_X$ and $\gamma_X$ their lifetime and Lorentz boost factor, respectively. Finally, $\brvis$ is the branching ratio of visible decays of particle $X$. Details of estimates are summarized in Appendix~\ref{app:snd-sensitivity-decays}. The sensitivities are shown in Fig.~\ref{fig:sensitivity-to-decays}, where we show the estimate for the Run 3 setup, as well as for the possible upgrade that may operate during Run 4. For the upgraded setup, the number of events at the lower bound is higher by a factor of $l_{\text{det,upgr}}^{\text{dec}}/l_{\text{det}}^{\text{dec}}\cdot \mathcal{L}_{\text{Run 4}}/\mathcal{L}_{\text{Run 3}} = 50$.

We conclude that for the Run 3 setup, SND@LHC may probe only a tiny parameter space for dark scalars, dark photons and HNLs that mix with $\tau$ flavor. For the upgraded setup, it may be possible to probe HNLs that mix exclusively with $\nu_{\mu}$ in the mass range $\lesssim \unit[2.5]{GeV}$, and in $\lesssim \unit[2.0]{GeV}$ for pure mixing with $\nu_\tau$. SND@LHC may also probe dark photons at the upper bound of the sensitivity with masses $m_V\lesssim\unit[0.1]{GeV}$, and dark scalars with masses $m_S\lesssim\unit[0.8]{GeV}$.

\section{Comparison with FASER}
\label{sec:discussion}
There is a similarity between the facilities of SND@LHC and FASER/FASER$\nu$ experiments. They are both placed in a large $\eta$ region and at the same distance $\lmin = 480$~m from the ATLAS interaction point, but in the opposite tunnels. Parameters of the experiments are summarized in Table~\ref{tab:snd-faser}. Below, we make a qualitative comparison of the sensitivities of the SND@LHC and FASER experiments, and then comment on the changes due to upgrades. 
\begin{table}[!h]
        \centering
        \begin{tabular}{|c|c|c|c|c|c|c|}
        \hline
        Detector & $l_{\text{min}},$~m & $l_{\text{det}},$~m & $\theta_{\text{min}}$, mrad & $\theta_{\text{max}}$, mrad &$\Omega \cdot 10^{7},\text{sr}$ & $\mathcal{L}, \unit{fb^{-1}}$\\
        \hline
        SND@LHC & \multirow{5}{*}{480} & 0.5 & 0.3 & 1.5 & $6.9$ &  \multirow{3}{*}{150} \\
        \cline{1-1} \cline{3-6}
             FASER&  & 1.5 & 0. & 0.2& $1.4$ &  \\
        \cline{1-1} \cline{3-6}
             FASER$\nu$ &  & 1. & 0. & 0.4&  $2.7$ & \\
             \cline{1-1} \cline{3-7}
             SND@LHC upgr. &  & 1.25 & 0.3 & 1.5&  $6.9$ &\multirow{2}{*}{3000}
        \\
        \cline{1-1} \cline{3-6}
             FASER2 &  & 5 & 0. & 2.1&  $138$ & \\
             \hline
        \end{tabular}
        \caption{Parameters of SND@LHC and FASER experiments: the distance to the decay volume, the length of the decay volume, the polar coverage, covered solid angle, total integrated luminosity.}
        \label{tab:snd-faser}
    \end{table}

Let us summarize the main differences between SND@LHC and FASER/FASER$\nu$ detectors in the reconstruction of signal. For scattering, SND@LHC competes with the FASER$\nu$ detector. FASER$\nu$ consists of emulsion films interleaved with tungsten plates, only providing the information of spatial position of different tracks with $30\%$ energy reconstruction accuracy for neutrino events (see also~\cite{Ismail:2020yqc}). For muons, the situation is much better, as they, being produced in FASER$\nu$, may penetrate it and enter FASER, which allows timing and momentum measurements. This option is unavailable, however, for other particles (hadrons, electrons), as they are effectively absorbed in the detector. On the contrary, SND@LHC provides timing measurements by the use of the SciFi technology and the energy reconstruction accuracy of $22\%$ for electrons. For both experiments, timing is needed for rejecting the background induced by high-energy muons and secondary particles.\footnote{In this work, we compare the signal of new physics to the number of neutrino interactions, which was already obtained under assumption of possible background. Therefore, these key features of the detectors' concept are omitted in our analysis.}

In the case of decays, SND@LHC competes with FASER, and their detectors provide comparable FIP parameters reconstruction accuracy, thanks to good spatial resolution of the emulsion. Assuming that SND@LHC is a background free experiment when searching for decays, the only relevant quantity for comparing the experiments is the number of correctly identified FIP decay events.

\subsection{Lower and upper bound of the sensitivity}
Let us now consider the differences in the number of events at these experiments. Two factors are important. First, SND@LHC is slightly off-axis, whereas FASER($\nu$) is placed directly on-axis. Second, SND@LHC covers $\simeq 5$ ($1.25$) times larger solid angle than FASER($\nu$). The different placements cause two effects that directly affect the lower and upper bound of the sensitivity (we follow~\cite{Bondarenko:2019yob} here). 

First, particles $X$ flying off-axis have smaller energies than those flying on-axis. This is important for probing FIPs that have small decay lengths $\ldec \lesssim \lmin$. Indeed, in this regime, the decay probability is $\pdec \approx \exp[-\lmin/c\tau_{X}\gamma_{X}]$. The sensitivity to such large couplings $g$ determines the upper bound, which is very sensitive to the mean energy of $X$:
\begin{equation}
\frac{g^{2}_{\text{upper,SND@LHC}}}{g^{2}_{\text{upper,FASER}}} \sim \frac{\gamma_{X}^{\text{SND@LHC}}}{\gamma_{X}^{\text{FASER}}}
    \label{eq:upper-bound}
\end{equation}
The upper bound is important for particles that may be probed by the FASER and SND@LHC experiments only in the regime of small decay lengths, including dark photons and axion-like particles (see Fig.~\ref{fig:sensitivity-to-decays}). The ratio of the mean $\gamma$-factors of dark photons $A'$, flying in the detector, is $\gamma_{A'}^{\text{SND@LHC}}/\gamma_{A'}^{\text{FASER}}\approx 1/3$. The resulting estimate~\eqref{eq:upper-bound} agrees with the sensitivities in Fig.~\ref{fig:sensitivity-to-decays}. 

Second, the off-axis placement may affect the geometric acceptance. Light portal particles $X$ are often produced in decays of mesons. The angular distribution of particles $X$ is similar to the distribution of parent mesons at angular scales larger than $\Delta \theta \simeq 2p_{X,\text{rest}}/\langle E_{\text{meson}}\rangle$, where $p_{X,\text{rest}}$ is the momentum of the daughter particle at rest frame of the decaying meson, being $\simeq m_{\text{meson}}$ if masses of all decays products are $\ll m_{\text{meson}}$. If $\Delta \theta > \theta_{\text{SND@LHC}} \simeq \mathcal{O}(\unit[1]{mrad})$, the ratio of geometric acceptances $\epsilon_{\text{geom}}$ for the SND@LHC and FASER experiments scales with their solid angle coverage. Using characteristic energies $\langle E_{\text{meson}}\rangle \simeq \unit[1]{TeV}$ for mesons produced in the far-forward region, we find that this scaling is indeed the case of light particles produced in decays of $D$, $B$-mesons. 

However, if the daughter particle is heavy $m_{X}\simeq m_{\text{meson}}$, or if the decaying meson is light (such as $\pi$, $\eta$, $K$), the geometric acceptance depends on the shape of the meson distribution. Experimental measurements of the meson production cross section in the region $|\eta| < 5$~\cite{Sirunyan:2017zmn,Aad:2015zix,Aaij:2013noa,Khachatryan:2016csy} provide the following scaling:
\begin{equation}
\label{eq:meson pT}
    \frac{\text{d}\sigma}{\text{d}p_\text{T}} \sim \frac{p_\text{T}}{(p_\text{T}^{2}+\Lambda_{\text{meson}}^{2})^{2}},
\end{equation}
independently of the pseudorapidity. The values of $\Lambda_{\text{meson}}$ are of order of $\Lambda_{\text{QCD}}\approx \unit[250]{MeV}$ for light mesons $\pi$, $\eta$, $K$, and $m_{D/B}$ for $D/B$-mesons. Numeric approaches (see, for instance,~\cite{Nason:1989zy,Pierog:2013ria,Riehn:2017mfm,Cacciari:2015fta}) predict the same behavior of $\text{d}\sigma/\text{d}p_\text{T}$ almost independently of pseudorapidity, including the far-forward region.\footnote{Some of these approaches suffer from theoretical uncertainties in far-forward direction~\cite{Bai:2020ukz}: small $p_\text{T}$ and large pseudorapidity require using parton distribution functions in the domain of small Bjorken scaling variable $x$, which are poorly constrained. One of the goal SND@LHC and FASER may serve for checking the distributions (and in particular the property~\eqref{eq:theta-max}) via studying the events with neutrinos produced in the meson decays.} This means that the meson distribution $\text{d}f/\text{d}\Omega$ is flat for angles $\theta \lesssim \theta_{\text{flat}}$, where
\begin{equation}
\theta_{\text{flat}}\sim \frac{\langle p_\text{T}\rangle}{\langle E_{\text{meson}}\rangle} \sim \frac{\Lambda_{\text{meson}}}{\unit[1]{GeV}}\unit{mrad} \simeq \begin{cases}
\mathcal{O}(\unit[1]{mrad}), \quad B,D \\ \mathcal{O}(\unit[0.1]{mrad}), \quad \pi,\eta\end{cases}
\label{eq:theta-max}
\end{equation}
Using the spectra of mesons (see Appendix~\ref{app:snd-sensitivity-decays}), we find
\begin{equation}
    \frac{\egeom^{\text{SND@LHC}}}{\egeom^{\text{FASER}}} \simeq \begin{cases} 1, \quad \pi,\eta, \\ \frac{\Omega_{\text{SND@LHC}}}{\Omega_{\text{FASER}}}\approx 5, \quad D, B,\tau\end{cases}
    \label{eq:egeom-ratio}
\end{equation}

\subsubsection{Decays}
Based on these findings, we can make a simple comparison of minimal couplings that may be probed by the FASER and SND@LHC experiments. Further, we will assume the most optimistic estimate for SND@LHC, according to which decays of FIPs may be clearly distinguished from backgrounds, and therefore, only $3$ events are required at 95\% CL.

In the regime $\ldec \gg \lmax$, the number of decay events of particles that originate from mesons is
\begin{equation}
    N_{\text{decay}} \propto \egeom \cdot \ldet \cdot g^{4}\cdot \brvis
\end{equation}
From this relation, we obtain
\begin{equation}
    \frac{g^{2}_{\text{lower,SND@LHC}}}{g^{2}_{\text{lower,FASER}}}
    \sim \sqrt{\frac{\gamma_{X}^{\text{SND@LHC}}}{\gamma_{X}^{\text{FASER}}}} \cdot \sqrt{\frac{\brvis^{\text{FASER}}}{\brvis^{\text{SND@LHC}}}}\cdot \begin{cases}1.7, &\text{particles from }\pi, \eta \\ 0.8, &\text{particles from $D$, $B$} \end{cases}
    \label{eq:snd-faser-sensitivity}
\end{equation}
where we used $\ldet^{\text{SND@LHC}} = \unit[0.5]{m}$. 

Comparing the lower bounds of the numerical sensitivities of SND@LHC and FASER for dark photons and dark scalars in Fig.~\ref{fig:sensitivity-to-decays}, we find that they agree with the estimates~\eqref{eq:snd-faser-sensitivity}. However, for HNLs there is a disagreement as large as a factor of $3$. A reason for this may be different distributions of $D$ mesons used in our analysis and in~\cite{Beacham:2019nyx} (see also Appendix~\ref{app:snd-sensitivity-decays}).

Let us now comment on the lower bounds ratio with the upgrade. With the help of the formulas~\eqref{eq:egeom-ratio},~\eqref{eq:snd-faser-sensitivity} and table~\ref{tab:snd-faser}, we conclude that the FASER2 experiment has much better potential: 
\begin{equation}
    \frac{g^{2}_{\text{lower,SND@LHC upgr}}}{g^{2}_{\text{lower,FASER2}}}\simeq 20\cdot \sqrt{\frac{\gamma_{X}^{\text{SND@LHC}}}{\gamma_{X}^{\text{FASER}}}} \cdot \sqrt{\frac{\brvis^{\text{FASER}}}{\brvis^{\text{SND@LHC}}}}
\end{equation}
A reason for this is mainly significantly larger angular coverage in the case of the FASER2.

 \subsubsection{Scattering}
Consider now the scattering signature. We will focus on the NC/CC, given the uncertain status of the background for the elastic signature. For the leptophobic portal, from Eq.~\eqref{eq:number-of-events-scattering}, the ratio of minimal probed couplings is (for $\alpha_\chi = \alpha_B$)
\begin{equation}
    \label{eq:lepto comparison1}
    \frac{\alpha_{B, \text{FASER}\nu}}{\alpha_{B, \text{SND@LHC}}} \sim  \left(\frac{\epsilon_{\text{geom}}^{\text{FASER}\nu}}{\epsilon_{\text{geom}}^{\text{SND@LHC}}} \frac{l^{\text{FASER}\nu}_{\text{det}}}{l^{\text{SND@LHC}}_{\text{det}}}
    \sqrt{\frac{N^{\text{FASER}\nu}_{\nu\text{ bg}}}{N^{\text{SND@LHC}}_{\nu\text{ bg}}}}\right)^{\frac{1}{3}}
\end{equation}
where $N_{\nu\text{ bg}}$ is the number of neutrino background events, and we assume that the detection efficiency is equal to one.

For small masses $m_V\lesssim 0.5$, the mediator is mainly produced from $\pi$, $\eta$ decays, as shown in Fig.~\ref{fig:V-prod}. In this case, we have $\epsilon^{\text{SND@LHC}}_{\text{geom}}/\epsilon^{\text{FASER}\nu}_{\text{geom}} \approx 0.3$. A similar increase occurs for $N_{\nu \text{ bg}}$, since neutrinos are abundantly produced in decays of pions, and therefore, we can use the same scaling for the total neutrino events, $N_{\nu \text{ bg}}\propto \epsilon_{\text{geom}} l_\text{det}$.
The estimate then reads:
\begin{equation}
    \label{eq:lepto comparison2}
    \frac{\alpha_{B, \text{SND@LHC}}}{\alpha_{B, \text{FASER}\nu}}  \simeq 1.5, \quad \text{   NC/CC signature}
\end{equation}
\section{Conclusions}
\label{sec:conclusions}
In this paper, we have demonstrated the potential of the SND@LHC experiment to probe feebly interacting particles. We have considered a few scattering and decay signatures. 

Light dark matter particles coupled via mediators may be searched by looking at the scattering signature, see Sec.~\ref{sec:scatterings}. These events need to be distinguished from neutrino scatterings. Because of large mass of $Z$ and $W$ bosons that mediate the neutrino interactions, the neutrino scattering occurs inelastically most of the times. This may be not the case for light dark matter particles interacting via a light $\mathcal{O}(\unit[1]{GeV})$ mediator, for which the yields of elastic and inelastic scattering events are comparable (see Fig.~\ref{fig:elastic-to-dis-ratio}). Therefore, looking for an excess in the yield of elastic scattering events may be suitable for probing such FIPs. For SND@LHC this, however, requires a dedicated study, since the search is performed using the emulsion data only, which is highly contaminated by the tracks from neutrino DIS events. For heavier mediators, FIPs scattering still may be searched via an increase in the ratio of scattering events with a lepton and those without a lepton. On one hand, this ratio may be accurately measured at SND@LHC. On the other hand, it is clearly predicted by the SM. We have illustrated the power of these two signatures by estimating the sensitivity to the scattering of light dark sector particles via the leptophobic portal, see Fig.~\ref{fig:leptophobic-sensitivity}.

SND@LHC detector may also search for decays of mediators, see Sec.~\ref{sec:decays}. Because of good spatial resolution of the emulsion in SND@LHC, decays into two charged particles may be distinguished from the neutrino scattering events. Such decays are main decay channel in the case of heavy neutral leptons, dark scalars that mix with Higgs boson, and dark photons. It is possible to probe their parameter space at its upgraded version as described in~\cite{TP}, see Fig.~\ref{fig:sensitivity-to-decays}. However, further studies of possible backgrounds are required to clarify these results.

We have also compared the potential of SND@LHC and FASER/FASER$\nu$ facilities to probe new physics, see Sec.~\ref{sec:discussion}. Placed at the same distance but at the opposite sides of the ATLAS experiment interaction point, they are very similar. There are a few factors, however, leading to differences in the sensitivity of these facilities to new physics. First, FASER is on-axis, while SND@LHC is slightly off-axis. The off-axis placement decreases the mean momentum of particles produced in the direction of SND@LHC, which somewhat worsens its potential to probe short-lived particles with the decay lengths of the order of the distance to the detector. Second, SND@LHC covers $\simeq 5$ times larger solid angle than FASER. Because of this, depending on the FIPs production channel, a fraction of FIPs flying in the direction of FASER is smaller than that for SND@LHC. For scatterings, FASER$\nu$ has higher event rate due to larger detector length and on-axis position, resulting in better sensitivity. This can be applied for the NC/CC signature.

\section*{Acknowledgements}
We thank M.~Cacciari, G.~De Lellis, M.~Ferrillo and F.~Kling for valuable discussions. This project has received funding from the European Research Council (ERC) under the European Union's Horizon 2020 research and innovation programme (GA 694896) and from the NWO Physics Vrij Programme “The Hidden Universe of Weakly Interacting Particles” with project number 680.92.18.03 (NWO Vrije Programma), which is (partly) financed by the Dutch Research Council (NWO).

\bibliographystyle{JHEP}
\bibliography{bibliography.bib}

\newpage

\appendix

\section{Scatterings vs decays}
\label{app:scattering-vs-decays}
Consider scatterings and decays in an example of vector (dark photons) and scalar portals. The scattering probability is
\begin{equation}
            P_{\text{scat}} = \sigma_{\text{scat}} n_{\text{atom}} L_{\text{det}}, \qquad n_{\text{atom}} \sim (\unit[1]{keV})^3
            \label{eq:scattering-prob}
        \end{equation}
        with the scattering cross section
        \begin{equation}
            \sigma_{\text{scat}} \sim \frac{\alpha_{S\chi\chi}y_{N}^{2}\theta^{2}}{m_{N}E_{\chi}}~(\text{scalar}),\qquad \sigma_{\text{scat}} \sim \frac{\alpha_D \epsilon^2}{m_V^2}~(\text{vector})
        \end{equation}
In its turn, the decay probability is
        \begin{equation}
            P_{\text{dec}} = \frac{\Gamma L_{\text{det}}}{\gamma},\qquad 
            \Gamma \sim \theta^2 \frac{m_S^3}{v^2}~(\text{scalar}),\qquad \Gamma \sim \epsilon^2 m_V~(\text{vector})
        \end{equation}
Comparing these two probabilities, one gets
        \begin{equation}
           \frac{P_{\text{scat}}}{P_{\text{dec}}} \sim \gamma \alpha_{S\chi\chi}y_{N}^{2} \left(\frac{v}{m_S}\right)^2 \frac{n_{\text{atom}}}{m_S m_{N}E_{\chi}}~(\text{scalar}),\qquad 
           \frac{P_{\text{scat}}}{P_{\text{dec}}} \sim \gamma \alpha_D \frac{n_{\text{atom}}}{m_V^3}~(\text{vector})
        \end{equation}

\section{Decay events}
\label{app:snd-sensitivity-decays}
We estimate the number of decays using the following formula:
\begin{equation}
    \nev = \sum_{\text{i}}N^{i}_{\text{prod}} \cdot \egeom^{i}\cdot \pdec^{i} \cdot \brvis,
\end{equation}
Here, $N_{\text{prod}}^{i}$ is the total number of particles $X$ produced via a channel $i$, $\egeom$ is the geometric acceptance, and \pdec is the decay probability averaged over energies of $X$, 
\begin{equation}
    \pdec^{i} = \int (e^{-\lmin/c\tau_{X}\gamma_{X}} - e^{-\lmax/c\tau_{X}\gamma_{X}})f^{i}_{E_{X}}dE_{X},
\end{equation}
Here, $\lmin = \unit[480]{m}$ is the distance to the SND@LHC detector, $\lmax - \lmin = \ldet^{\text{decay}}$, $f_{E_{X}}$ is the energy distribution of particles $X$ that fly in the decay volume. Finally, $\brvis$ is the branching ratio of visible decays.

HNLs that mix with $\nu_{\mu}$ are produced in decays of $D_{s}/D^{+/0}$ mesons. HNLs that mix with $\nu_{\tau}$ are produced mainly in decays of $\tau$-leptons, which, in their turn, originate from decays $D_{s}\to \tau \bar{\nu}_{\tau}$~\cite{Bondarenko:2018ptm}. We have obtained the distribution of $D$ mesons using SIBYLL 2.3c~\cite{Riehn:2017mfm,Fedynitch:2018cbl} as a part of the CRMC package~\cite{CRMC}. As a cross-check, for the charm production we have compared the predictions of SIBYLL with results of the FONLL program~\cite{Cacciari:1998it,Cacciari:2012ny,Cacciari:2015fta}. We have found that the results agree well for angles $\theta > 0.8\text{ mrad}$.\footnote{For smaller angles, FONLL (both the online form and installed program) predicts zero or negative cross sections, which indicates some internal problem.} Having the $D$ distribution, we have obtained the distribution of $\tau$-leptons and, subsequently, HNLs angles and momenta using the approach described in~\cite{Boiarska:2019vid}. For simplicity, we approximate the angle-momentum distribution of HNLs by that of particles produced in a two-body decay $\tau \to \pi N$ (for the mixing with $\nu_{\tau}$) and $D_{s}\to \mu  N$ (for the mixing with $\nu_{\mu}$). 

Dark photons $V$ in sub-GeV mass range are produced in decays $h = \gamma V$ of $\pi$- and $\eta$-mesons, and by proton bremsstrahlung~\cite{deNiverville:2016rqh}. We use the angle-energy distributions of the mesons generated by EPOS-LHC~\cite{Pierog:2013ria} as a part of the CRMC package~\cite{CRMC}, and follow~\cite{deNiverville:2016rqh} for the bremsstrahlung.

Dark scalars $S$ are produced in decays $B\to X_{s}S$ of $B$-mesons, where $X_{s}$ is a hadron including an $s$-quark, and by the proton bremsstrahlung~\cite{Boiarska:2019jym}. We use FONLL in order to obtain the angle-energy distribution of $B$-mesons, and follow~\cite{Boiarska:2019jym} for the proton bremsstrahlung. 

\begin{figure}[!h]
    \centering
    \includegraphics[width=0.5\textwidth]{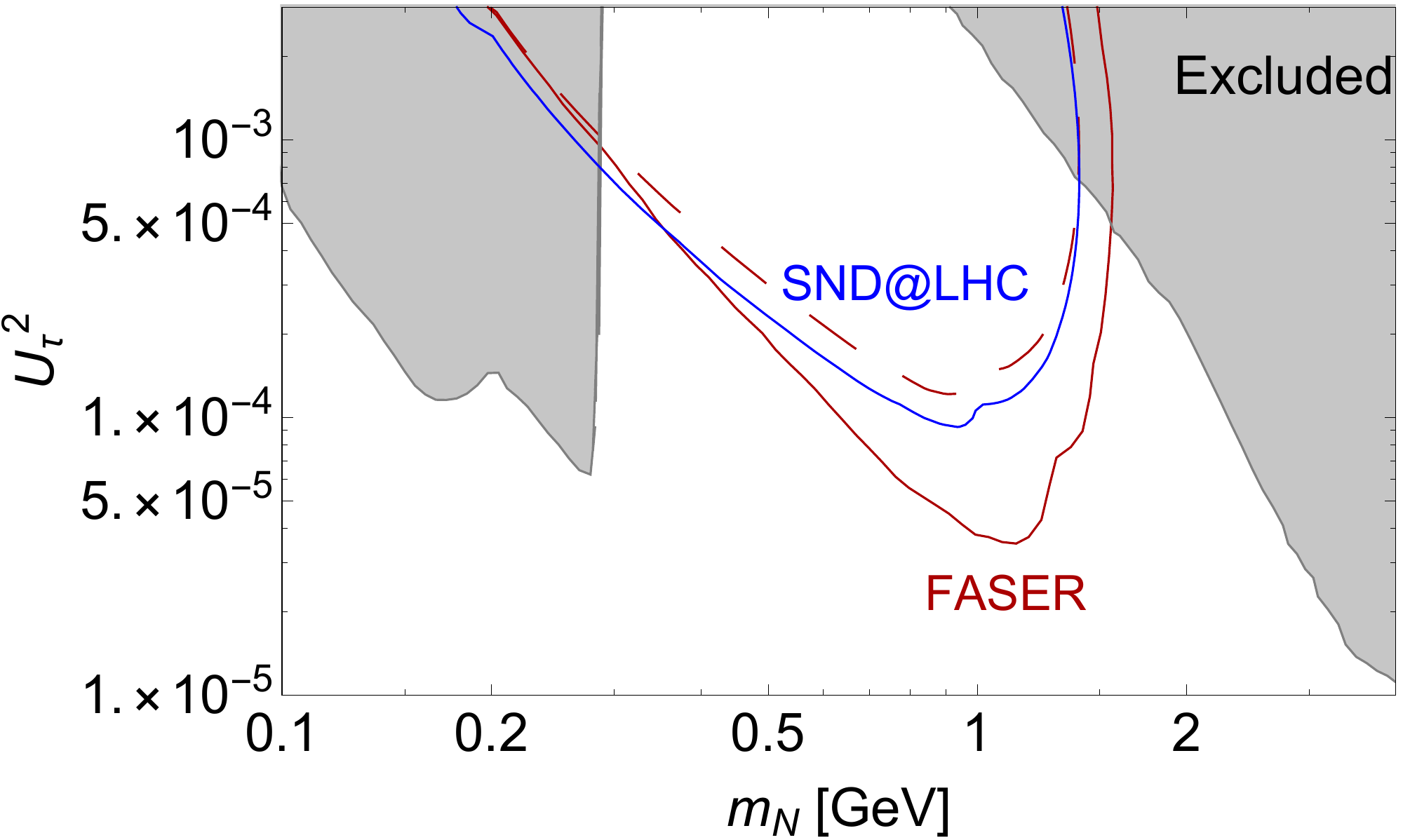}
    \caption{The sensitivity of FASER to HNLs that mix with $\nu_{\tau}$. The solid line corresponds to the contour given in Ref.~\cite{Beacham:2019nyx}, while the dashed line -- to our estimate. For the comparison, we also show the sensitivity of SND@LHC (in blue).}
    \label{fig:faser-HNLs}
\end{figure}

Using the obtained distribution, we have reproduced the sensitivity of FASER to scalars and dark photons from~\cite{Beacham:2019nyx}. However, we have not reproduced the sensitivity to HNLs, see Fig.~\ref{fig:faser-HNLs}. A reason may be in different distributions of $D_{s}$-mesons used in the estimates.

\section{Leptophobic mediator: production and decays}
\label{app:leptophobic-phenomenology}
In order to describe interactions of $V$ with hadrons, we follow~\cite{Fujiwara:1984mp} (see also~\cite{Tulin:2014tya}), in which vector mesons $m$ play the role of gauge bosons of a ``hidden'' local $SU_{f}(3)$ symmetry in the space of pseudoscalar mesons nonet. The EM field is included as a background field that is associated with the appropriate generator $Q = \text{diag}\left(\frac{2}{3},-\frac{1}{3},-\frac{1}{3}\right)$, and mix with the vector mesons. The coupling of the vector mesons to the pseudoscalar mesons is fixed by the anomalous decay $\pi^{0}\to \gamma\gamma$. This model is very successful in describing the EM scattering data $e^{+}e^{-}\to \text{hadrons}$ and decay widths of vector mesons. We assume that it may be also used for describing the phenomenology of the leptophobic boson.

For the lephophobic mediator, the generator is $T_{V} = \frac{\mathbb{1}}{3}$, and its mixing coupling is given by
\begin{equation}
    f_{Vm} = -2g_{B}g_{m}\text{Tr}[T_{V}T_{m}],
\end{equation}
where $T_{m}$ is a generator associated with the given meson, and $g_{m}/m_{m}^{2} = 1/\sqrt{12\pi}$, as fixed by the anomaly. The mixing occurs only with isosinglet $\omega$- and $\phi$-mesons, for which
\begin{equation}
    T_{\omega} = \frac{1}{2}\text{diag}(1,1,0), \quad T_{\phi} = \frac{1}{\sqrt{2}}\text{diag}(0,0,1)
\end{equation}
The decay width of $V$ may be extracted from the experimental data on the EM ratios $\sigma(e^{+}e^{-}\to \text{hadrons})/\sigma(e^{+}e^{-}\to \mu^{+}\mu^{-})$, where the hadronic final states correspond to $\phi$-like and $\omega$-like decays. This has been made in~\cite{Ilten:2018crw}, in which the data have been used for describing the decay widths up to masses $m_{V}\simeq \unit[1.7]{GeV}$, while for larger masses perturbative calculations were used. We use the results of this paper. 

The resonant enhancement is also important when considering the production of the mediator by the proton bremsstrahlung by affecting the form-factor $F_{ppV}$ in the $ppV$ vertex.  The baryonic form factor $F_{ppV}$ may be related to the proton and neutron EM dipole form-factors $F_{p/n}$, which are, in its turn, related to the isoscalar form factor $F_\omega \equiv \frac{F_p + F_n}{2}$, which in the extended vector meson dominance model coincides with the $\omega$ contribution~\cite{Faessler:2009tn}:\footnote{We assume no contribution of the $\phi$-meson to the form-factor, since the corresponding coupling $f_{\phi NN}$ is expected to be suppressed~\cite{Kuwabara:1995ms,Nakayama:1999jx} (i.e., neglecting the $s$-quark contribution in the proton PDF).}
\begin{equation}
    \langle p | J_B|p\rangle = \langle p|J_{\text{EM}} |p\rangle + \langle n|J_{\text{EM}}|n\rangle\longrightarrow F_{ppV} = 2F_\omega
\end{equation}
Unfortunately, the experimental data on $e^{+}e^{-}\to p^{+}p^{-}$, which may be used for extracting the EM form-factors in the time-like region, is limited by the physical threshold $q^{2}>4m_{p}^{2}$. Following~\cite{Faessler:2009tn} (see also~\cite{Kling:2020iar}), for extrapolating in the domain of lower invariant masses we use
\begin{equation}
    F_{ppV}(q^{2}) = \sum_{ \omega}f_\omega\frac{m_{\omega}^{2}}{m_{\omega}^{2}-q^{2}-i\Gamma_{\omega}m_{\omega}},
    \label{eq:form-factor}
\end{equation}
where the sum goes over $\omega(782), \omega(1420),\omega(1680)$, $f_\omega= 2 f_{NN
\omega}/g_{\omega}$, with $f_{NN\omega}$ being the meson's coupling to the nucleon, while $g_{\omega}$ is the meson's coupling to photon. We use the couplings $f_{NN\omega(782)} = 17.2$ and $g_{\omega(782)} = 17.1$~\cite{Faessler:2009tn}. 
The couplings to the other two resonances are unknown. However, the remaining two coefficients $f_{\omega(1420)} = -2.16+0.77i$ and $f_{\omega(1680)} = 1.14-0.57i$ in Eq.~\eqref{eq:form-factor} may be fixed by two requirements: $F_{ppV}(0) = 1$, and $F_{ppV}(-q^{2}) \sim 1/q^{4}$. The first requirement comes from the fact that the form-factor $F_{ppV}$ is reduced to the baryon charge at low momenta transfer. The second requirement comes from the behavior of the proton's dipole form-factor in the space-like region predicted by the quark counting rules~\cite{Vainshtein:1977db}.

\begin{figure}[!h]
    \centering
    \includegraphics[width=0.5\textwidth]{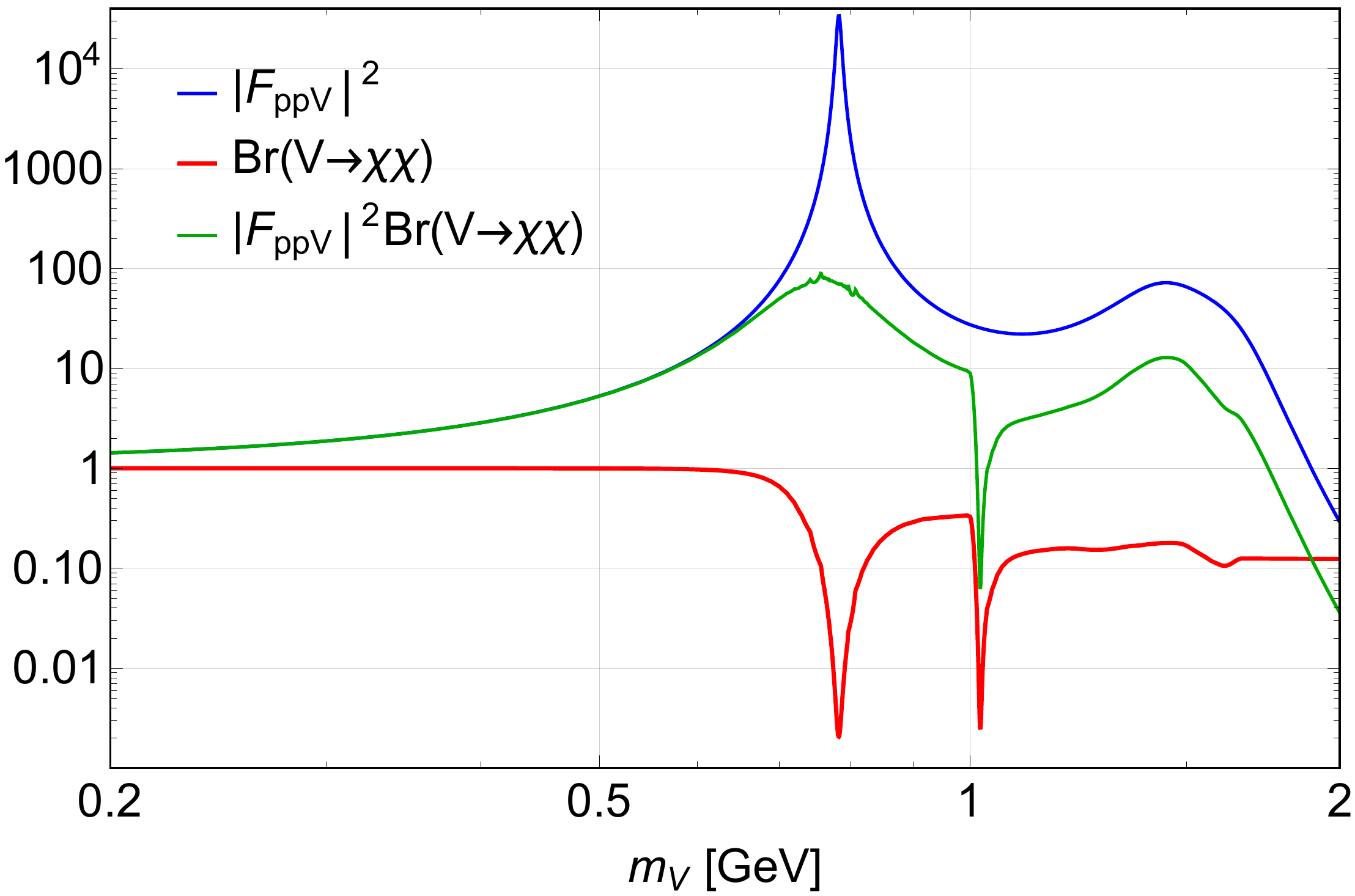}
    \caption{The behavior of the $ppV$ form-factor~\eqref{eq:form-factor} and the branching ratio for the process $V\to \chi\chi$. The coupling $\alpha_{\chi} = \alpha_{B}$ is assumed, and $m_\chi = m_V/3$.}
    \label{fig:form-factor-behavior}
\end{figure}

The behavior of the branching ratio into a $\chi\chi$ pair and the form-factor is shown in Fig.~\ref{fig:form-factor-behavior}. Note that for the choice $\alpha_{\chi} = \alpha_{B}$, commonly considered in the literature, the enhancement of $F_{ppV}$ near $m_{V} = m_{\omega(770)}$ and suppression of $\text{Br}(V\rightarrow \chi\chi)$ due to the $\omega$ resonances cancel each other.

\section{Elastic and inelastic scattering cross sections for the leptophobic portal}
\label{app:scattering-cross sections}
\subsection{Elastic scattering}
The cross section of the elastic scattering is
\begin{equation}
    \sigma_{\text{elastic}} = \int dE_{\chi}f_{E_{\chi}} \int \limits_{E_{N,\text{min}}}^{E_{N,\text{max}}(E_{\chi})}dE_{N}\frac{d\sigma_{\chi N \to 
    \chi N}}{dE_{N}},
\end{equation}
where $E_{N,\text{min}}$ is the minimal recoil energy that may be detected, the maximal recoil energy of the nucleon is
\begin{equation}
E_{N,\text{max}} = \frac{m_{N} (2 E_{\chi}^{2} + 
      2 E_{\chi} m_{N} + m_{N}^{2} - 
      m_{\chi}^{2})}{2 E_{\chi} m_{N} + m_{N}^{2} +
    m_{\chi}^{2}},
\end{equation}
$Q^{2} = 2m_{N}(E_{N}-m_{N})$ is the modulus of the squared momentum transfer, $Q^{2} = -(p_{\chi}-p_{\chi}')^{2}$. Finally, the differential cross section is
\begin{equation}
    \frac{d\sigma_{\chi N \to 
    \chi N}}{dE_{N}} =4\pi \alpha_{D}^{2} F_{N}(Q^{2}) \frac {m_{N} (2 E_{\chi}^{2} + 2 E_{\chi} m_{N} + m_{\chi}^{2}) - 
   E_{N} (2 E_{\chi} m_{N} + m_{\chi}^{2})}{(E_{\chi}^{2} - 
     m_{\chi}^{2}) (2 E_{N} m_{N} - 2 m_{N}^{2} + m_{V}^{2})^{2}},
\end{equation}
where $F_{N}(Q^{2})$ is the elastic form-factor, which we assume to be $F_{N}(Q^{2}) = \frac{1}{(1+\frac{Q^2}{\unit[0.71]{GeV^2}})}$.
\subsection{Inelastic scattering}
In the case of the inelastic scattering, we follow~\cite{Soper:2014ska}, which uses the parton model. Let us introduce the variables $E_{V} = E_{\chi}-E_{\chi}^{'}$, $Q^{2}$. The differential cross section is
\begin{equation}
    \frac{d^{2}\sigma}{dE_{V}dQ^{2}} =\frac{\pi\alpha_{D}^{2}}{9m_{N}}\frac{1}{E_{\chi}^{2}-m_{\chi}^{2}}\frac{1}{(m_{V}^{2}+Q^{2})^{2}}(2p-q)^{\mu}(2p-q)^{\nu}W_{\mu\nu}\sum_{q}xf_{q}(x,Q^{2}),
\end{equation}
where $f_{q}(x,Q^{2})$ is the parton distribution function ($q = u/\bar{u}/d/\bar{d}/s/\bar{s}$), $x = \frac{Q^{2}}{2m_{N}E_{V}}$, $W_{\mu\nu}$ is the hadronic tensor,
\begin{equation}
    W_{\mu\nu} = -g_{\mu\nu} + \frac{q_{\mu}q_{\nu}}{q^{2}} + \frac{2x}{p_{N}\cdot q + 2x m_{N}^{2}}\left(p_{N\mu} - \frac{p_{N}\cdot q}{q^{2}}q_{\mu}\right)\left(p_{N\nu} - \frac{p_{N}\cdot q}{q^{2}}q_{\nu}\right)
\end{equation} 
Because of the property $q^{\mu}W_{\mu\nu} = q^{\nu}W_{\mu\nu} = 0$, we have
\begin{equation}
    (2p-q)^{\mu}(2p-q)^{\nu}W_{\mu\nu} =\frac{4 E_{\chi}^{2} Q^{2} - 4 E_{V} E_{\chi} Q^{2} - 
    Q^{4}}{E_{V}^{2} + Q^{2}} - 4 m_{\chi}^{2}
\end{equation}
The kinematic limits are
\begin{equation}
    Q^{2} < 2m_{N}E_{V}, \quad 2\mu^{2}<Q^{2}<4(E_{\chi}(E_{\chi}-E_{V})-m_{\chi}^{2})-2\mu^{2},
\end{equation}
\begin{equation}
    E_{\text{min}}<E_{V}<\frac{2m_{N}(E_{\chi}^{2}-m_{\chi}^{2})}{2E_{\chi}m_{N}+m_{N}^{2}+m_{\chi}^{2}},
\end{equation}
where $E_{\text{min}}$ is the minimal recoil, and the function $\mu$ is
\begin{equation}
    \mu^{2} =\frac{m_{\chi}^{2}E_{V}^{2}}{E_{\chi}(E_{\chi}-E_{V})-m_{\chi}^{2}-\sqrt{(E_{\chi}(E_{\chi}-E_{V})-m_{\chi}^{2})^{2}-m_{\chi}^{2}E_{V}^{2}}} 
\end{equation}
To get $f_{q}(x,Q^{2})$, we use LHAPDF with CT10nlo PDF sets. We assume that they are zero if $Q < \unit[1]{GeV}$.

\end{document}